\PassOptionsToPackage{pdfpagelabels=false}{hyperref} 

\documentclass[a4paper,fleqn,usenatbib]{mnras}

\usepackage{txfonts}

\usepackage[T1]{fontenc}
\usepackage{ae,aecompl}

\usepackage{color}

\usepackage{graphicx}
\usepackage{amstext}

\usepackage{hyperref}
\usepackage{bookmark}
\usepackage{booktabs, caption, makecell}
\usepackage{threeparttable}
\usepackage{relsize}
\usepackage{dirtytalk}
\usepackage{tipa}
\usepackage{etoolbox}
\makeatletter
\patchcmd\@combinedblfloats{\box\@outputbox}{\unvbox\@outputbox}{}{%
   \errmessage{\noexpand\@combinedblfloats could not be patched}%
}%
 \makeatother
\newcommand{\dm}{$\dot{M}$}

\newcommand\Tstrut{\rule{0pt}{2.6ex}}         % = `top' strut
\newcommand\Bstrut{\rule[-0.9ex]{0pt}{0pt}}   % = `bottom' strut

\definecolor{darkred}{rgb}{0.55, 0.0, 0.0}

\title[State transition luminosities]{Investigating state transition luminosities of Galactic black hole transients in the outburst decay}

\author[A. Vahdat Motlagh et al.]{
A. Vahdat Motlagh$^{1,2}$\thanks{E-mail: Armin.Vahdat@ttu.edu}, E. Kalemci$^{3}$, T.~J. Maccarone$^{1}$%, T. Din\c{c}er$^{5}$, T.~D. Russell$^{6}$
\\
$^{1}${Department of Physics and Astronomy, Texas Tech University, Box 41051, Lubbock, TX 79409-1051, USA}\\
$^{2}${Istanbul Technical University, Faculty of Science and Letters, Physics Engineering Department, 34469, Istanbul, Turkey}\\
$^{3}${Faculty of Engineering and Natural Sciences, Sabanc\i\ University, Orhanl\i-Tuzla, 34956, Istanbul, Turkey}\\
}

\date{Accepted XXX. Received YYY; in original form ZZZ}

\pubyear{2018}
\begin{document}
\label{firstpage}
\pagerange{\pageref{firstpage}--\pageref{lastpage}}
\maketitle
%\begin{document}
%\date{submitted to mnras}
%\pagerange{\pageref{firstpage}--\pageref{lastpage}} \pubyear{2014}
%\maketitle

%% /*******************************************************************
%% ** The Abstract                                                   **
%% *******************************************************************/

\begin{abstract}
We have performed a comprehensive spectral and timing analyses of Galactic black hole transients (GBHTs) during outburst decay in order to obtain the distribution of state transition luminosities. Using the archival data of the Rossi X-ray Timing Explorer (RXTE), we have calculated the weighted mean for state transition luminosities of 11 BH sources in 19 different outbursts and for disk and power-law luminosities separately. We also produced histograms of these luminosities in terms of Eddington luminosity fraction (ELF) and fitted them with a Gaussian. Our results show the tightest clustering in bolometric power-law luminosity with a mean logarithmic ELF of -1.70 $\pm$ 0.21 during the index transition (as the photon index starts to decrease towards the hard state). We obtained mean logarithmic ELF of -1.80 $\pm$ 0.25 during the transition to the hard state (as the photon index reaches the lowest value) and -1.50 $\pm$ 0.32 for disk blackbody luminosity (DBB) during the transition to the hard-intermediate state (HIMS). We discussed the reasons for clustering and possible explanations for sources that show a transition luminosity significantly below or above the general trends.
 
\end{abstract}

\begin{keywords}
stars: black holes - X-rays: binaries - accretion, accretion discs
\end{keywords}

%% /*******************************************************************
%% ** Introduction                                                   **
%% *******************************************************************/

\section{Introduction}\label{sec:intro}
Galactic black hole transients (GBHTs) are systems with a low-mass optical companion which spend most of their time in a faint quiescent state where almost no activity is observed. Occasionally, they undergo sudden and bright X-ray outbursts which usually last from a few weeks to a few months. However, in some sources like GRS 1915+105, it may take up to decades. During this period, X-ray flux increases by several orders of magnitude. Currently, the standard approach to understand the global evolution of a GBHT during the outburst is to examine its hardness-intensity diagram \citep[HID; see e.g.,][]{Homan2001ApJS}, in which the intensity (X-ray luminosity or count rate) is plotted against the hardness ratio (HR). GBHTs generally follow an anti-clockwise, \textit{q} shape track on the HID diagram and the changes in the  timing properties of the sources are strongly correlated with their spectral changes. Being completely model independent, the HID is also helpful in studying the spectral states. The current classification of spectral states based on the observational features are discussed in \cite{McClintock06book} and \cite{Belloni10_jp}. At the beginning of the outburst, the source is mainly characterized by a hard energy spectrum dominated by a power-law component ($\Gamma \approx 1.5$) and strong aperiodic variability (rms $\approx$ 30\%). This is known as the hard state (HS). As the power-law softens, aperiodic variability drops off (rms $\approx$ 1\%) and a quasi-thermal disc (with a typical characteristic temperature of ${k_B}T \approx$ 1 keV) along with a strong reflection component dominates the spectra (softstate, SS). In the classification scheme of \cite{Belloni10_jp}, temporal properties such as the rms variability, the shape of the power spectrum and the type of QPOs have also been utilized in order to distinguish hard and soft intermediate states (HIMS and SIMS). The transition between soft and hard states occurs at different luminosities in the rise and decay which is known as the hysteresis loop in the $q$ diagram \citep{Miyamoto1995ApJ}. The  loop does not behave in an identical manner for all sources in all outbursts and especially the upper region of the $q$ diagram can be very complicated.

Historically, states transition have been linked to changes in the mass accretion rate ($\dot{M}$). The hard state can be described by a two-component model where a geometrically thin and optically thin cool disk \citep{Shakura1973A&A} is truncated due to evaporation at some radius larger than the innermost stable circular orbit, ISCO,\citep{esin1997advection}. Inside, the matter is transported via a geometrically thick and optically thin accretion flow referred to as hot flow or corona. The truncation radius is comparably large, and although the geometry of the hot flow is still in debate, it is generally accepted that the observed power-law spectra originate from inverse Compton scattering of cool disk photons \citep[or possibly cyclo-synchrotron photons at low luminosities,][]{Sobolewska2011MNRAS, Poutanen14, 2013MNRASSkipper}. As $\dot{M}$ increases, the cool disk starts to move radially inward and the hot flow cools off rapidly by the low energetic seed photons and eventually collapses to the point where the energy spectrum is completely dominated by the disk. At the decay phase on the other hand, the corona reforms  gradually over time (possibly as the disk recedes) and the outburst cycle completes when the source returns back to its quiescence state. GBHTs also display complicated multi-wavelength properties during the outburst cycle which could provide additional information about the accretion geometry. For instance, radio and optical-infrared (OIR) measurements usually point out the presence of a compact jet which shows a flat to inverted radio spectrum in the HS \citep{kalemci2013complete}.

Frequent monitoring of the black holes with the Rossi X-ray Timing Explorer (RXTE) satellite made it clear that the observed transitions between states cannot be explained only with changes in \dm, and an additional parameter should be involved in transitions \citep{Homan2001ApJS}. Some of the alternative models for explaining state transitions are based on magnetic field generation and transport \citep{Petrucci2008MNRAS,Begelman2014ApJ,yan2015parameter,contopoulos2015cosmic,Kylafis2015A&A}, or general relativistic (GR) Lense \& Thirring effect \citep{nixon2014physical}.

%% /*******************************************************************
%% ** State transition luminosities                                  **
%% *******************************************************************/

\subsection{State transition luminosities}
\label{stl}

Although the nature of state transitions have been studied for more than four decades, the physical origin has not been well understood. Knowing what physical processes dominate during a state transition may hold a key to understand the accretion environment and the radiation mechanism during different spectral states. One way to approach this problem is to quantify the state transition luminosity clustering in transition to the hard state and study the impact of different observables (e.g. inclination, spectral model parameters) in the distribution of state transition luminosities.  For example, if a clear dependence on the inclination angle is seen, this would indicate that the emission is strongly anisotropic.  It is expected that the thermal components should come from geometrically thin, optically thick disks that should show fluxes proportional to the cosine of the inclination angle.  On the other hand, the hard X-ray components come from optically thin regions, and, for them, strong inclination angle dependences would most likely indicate either geometric beaming \citep[e.g.,][]{Beloborodov2001MNRAS, Markoff2001A} or perhaps seed photons dominated by the geometrically thin disk rather than cyclo-synchrotron radiation from within the hot flow itself \citep{Sunyaev1980A}.

 In this regard, the first attempt was made by \cite{maccarone2003x} where they studied the transition luminosities for 10 individual sources including both BHs and NSs with well-determined mass and distance. They investigated the sources in the outburst decay (soft-to-hard transition) and found the average state transition luminosity to be 1.9 $\pm$ 0.2  \% Eddington. 
 
A larger sample of transient and persistent black hole sources including the ones with poor (or non-existent) mass and distance estimates were studied by \cite{dunn2010global} in which the hardness intensity/luminosity diagrams (HID/HLD) were compared and discussed using the \textit{RXTE} data. The disk/power-law fractional luminosities of 13 GBHTs have been calculated for the outburst rise and decay. Hard to soft state transition luminosities are found -0.51 $\pm$ 0.41 in terms of logarithmic Eddington Luminosity Fraction (ELF) (30.9 $\pm$ 29.2 ELF) whereas soft to hard state transition luminosities are obtained -1.57 $\pm$ 0.59 log ELF (2.69 $\pm$ 3.65 ELF). \cite{Tetarenko2016ApJS} used a larger set of black hole sources and instruments and a statistical approach to study the state transition luminosity distributions and found an average transition luminosity value similar to that of \cite{dunn2010global} and \cite{maccarone2003x} for the decay part, but with a significantly lower value of 11.5 ELF for the rise, probably because \cite{Tetarenko2016ApJS} used a sample including outbursts not just from RXTE, but also from other missions with more sensitive all-sky monitors which allowed the detection of outbursts with fainter peaks.

Further quantifying the state transition luminosity distribution during the outburst decay is important for two key reasons. 
First, if the distribution is confirmed to be narrow as has been previously found, this provides support for the idea that a transition luminosity  depends strongly on only a single parameter, which does not vary much from source to source or from outburst to outburst.  It has often been suggested that the primary parameter for the state transition luminosity is the dimensionless viscosity parameter $\alpha$ \citep[e.g.][]{Narayan1994ApJ, Zdziarski1999MNRAS}, so the breadth of the state transition luminosity distribution may provide insights about how much $\alpha$ varies from source to source.  Second, given the prior results that indicate a relatively narrow state transition luminosity distribution, it has been suggested that the state transitions can be used as standard candles to estimate the distances to sources, especially in cases where the sources are highly extinct, so that other distance estimation techniques are ineffective  \citep[e.g.][]{maccarone2003x,Homan2006MNRAS, Jones2012M, 2015MNRASRussell}.  A further, more detailed quantification of the state transition luminosity distribution will help understand the level of precision that can be obtained using this method.
.

In this study, we have determined the state transition luminosities of GBHTs in the outburst decay by using the state definitions in \cite{Belloni10_jp} and \cite{kalemci2013complete}. We classified each observation in a single state based on these definitions and took the dates for which the states change as the date of state transition. Unlike the previous studies, we took account both spectral and temporal changes accordingly in order to determine the states and transition times (which might be quite challenging in intermediate states). Furthermore, the disk and power-law luminosities were obtained separately according to the procedure in \S\ref{sec:spect}. Our analyses cover 11 GBHTs which went through 19 outbursts in total. 

%% /*******************************************************************
%% ** Observations and Analysis                                      **
%% *******************************************************************/

\section{Observations and Analysis}\label{sec:obs}

\subsection{Data reduction and spectral analysis}\label{sec:spect}

The \textit{RXTE} data reduction was done with HEASoft (version: 6.19.2). Data were accumulated only when the spacecraft was pointing more than 10$^{\circ}$ above the horizon (elevation degree). Furthermore, data were rejected for 30-minute intervals in an orbit beginning with the satellite entering the South Atlantic Anomaly (SAA) in order to prevent possible contamination from activation in the detectors due to high energy particles in the SAA region. The time intervals with strong electron flares were also removed.

We used both the PCA and the HEXTE data whenever both were available and used only the PCA data if HEXTE was not available. After the spectral extraction, we added 0.8\% up to 15 keV and 0.4\% above 15 keV as systematic errors to the PCA spectra based on fits to Crab observations \citep{jahoda2006calibration}. We fitted the PCA data in the 3-25 keV energy band using the response matrix and the background model generated with the standard FTOOL programs\footnote{See \url{https://heasarc.nasa.gov/ftools/ftools\_menu.html} for more
detailed information.}. In order to maximize the number of counts, we used all available PCUs. For HEXTE, we used $\sim$15-200 keV energy range. When both clusters were available, we combined Cluster A and B spectra after making sure that the background regions of both clusters are free of contaminating sources\footnote{\url{https://heasarc.gsfc.nasa.gov/docs/xte/whatsnew/big.html}}. Extraction of the response matrix, background spectrum, and dead-time correction was done following the procedures described in the RXTE Cookbook\footnote{\url{http://heasarc.gsfc.nasa.gov/docs/xte/recipes/cook\_book.html}}.

We used the XSPEC package (version: 12.9.0n) for spectral fitting \citep{1996ASPC}. We set solar abundances to \textit{wilm} \citep{wilms2000absorption} and cross-section table to \textit{vern}  \citep{verner1996atomic} in XSPEC. We started our spectral fit with a combination of a multicolor disk-blackbody (``\textit{diskbb}''), a power law (``\textit{power}''), an absorption model (``\textit{tbabs}'') and a smeared edge (``\textit{smedge}''). Then, for each observation, we performed an F-test in order to determine if an iron line (emission around 6.4 keV,``\textit{gauss}'') \citep{ebisawa1994spectral} needs to be added. Similary, if required, we added an exponential cut-off (``\textit{hecut}'') in the power-law predicted by thermal Componization models. We added the new components only when F-test gave a probability less than 0.5\%. We emphasize that we did not use F-test to justify presence of iron lines \citep{Protassov2002ApJ}, but just to get a better overall fit to the spectrum which resulted up to 5\% variations in PL flux and up to 15\% variations in DBB flux.

%% /*******************************************************************
%% ** Temporal analyses                                              **
%% *******************************************************************/

\subsection{Temporal analyses}\label{sec:temp}

Temporal analyses were applied to each observation in order to extract timing information such as the rms amplitude of variability and the type of the quasi-periodic oscillations (QPO). Such information has provided a more precise distinction between states.

For the whole data set, we computed the PSD from the PCA data using the ``T\"{u}bingen timing tools '' in $3-25$ keV energy range. The power spectra were normalized as explained in \cite{Miyamoto1989} and \cite{Belloni1990}. This method not only allows taking the background into account, but also provides a much clear comparison of systematic brightness-independent similarities between different PSDs. The dead-time correction was done according to \cite{Zhang1995ApJ} using 10 $\mu$s dead-time per event.

The power spectra were fitted with a combination of broad and narrow Lorentzians \citep{Belloni2002ApJ} which allowed us to find the total rms amplitude of variability, as well as to classify QPOs.

\begin{table*}
\centering
\caption{Observational parameters of GBHTs used in this study.}
\label{tab:bhtransients}
\footnotesize
\begin{tabular}{l c c c c c c c c }
 \hline \hline 
                Source       		& & %1
                Mass            &
                Dist.         		&  % 3
                   %4
                Inclination                       &  %5
                Binary		&  %6
                Binary$ ^{\rm ^{\scalebox{0.96}{{\color{blue}a}}}}$ 	&%7
                N$_{\rm H}$ & 
                References$ ^{\rm ^{\scalebox{0.96}{{\color{blue}b}}}}$\\
	      	 			&  %1
	      			& %2
	      		(M$_{\odot}$)
	      	         	&
	      			(kpc)  & %3
	      		 %4
	      ($^{ \circ}$)	 & %5
	      period (h)             &  %6
	      sep.               & %7
	      (10$^{22}$)                 &
	      \\
	      
\hline 
\\

\multicolumn{2}{l}{4U 1543$-$47} & 9.4 ${\color{darkred}\pm}$ 2.0   & 7.5 ${\color{darkred}\pm}$ 0.5  & 20.7 ${\color{darkred}\pm}$ 1.5 $^{\rm ^{\scalebox{0.99}{$\dagger$}}}$ &   26.8 & 23 & 0.43 & 1, 2, 3 \\ %1, 2, 3, 3, 0, 1

\multicolumn{2}{l}{GRO J1655$-$40}$^{ \hspace{-3mm}\rm ^{\scalebox{0.96}{{\color{blue}c}}}}$ & 5.4 ${\color{darkred}\pm}$ 0.3  & 3.2 ${\color{darkred}\pm}$ 0.2     & 70.2 ${\color{darkred}\pm}$ 1.2 $^{\rm ^{\scalebox{0.99}{$\dagger$}}}$   & 62.4 ${\color{darkred}\pm}$ 0.6   & 38 & 0.8 & 4, 5, 6  \\  %4, 4, 5, 5, 0, 6

   \multicolumn{2}{l}{GX 339$-$4} & 9.0 ${\color{darkred}\pm}$ 1.4   &  8.4 ${\color{darkred}\pm}$ 0.9   & < 60 $^{\rm ^{\scalebox{0.99}{$\dagger$}}}$  &     42.2 & 25 & 0.57 & 7, 8, 9, 10   \\ %7, 7, 8, 9, 10, 0

\multicolumn{2}{l}{H1743$-$322} &  {\color{blue}8} ${\color{darkred}\pm}$ {\color{blue}1.5}  &  8.5 ${\color{darkred}\pm}$ 0.8   & 75 ${\color{darkred}\pm}$ 3 $^{\rm ^{\scalebox{0.99}{$\ddagger$}}}$  & ... & ...  & 2.3 & 11, 12   \\

\multicolumn{2}{l}{XTE J1550$-$564}  &  9.1 ${\color{darkred}\pm}$ 0.6  &  4.4 ${\color{darkred}\pm}$ 0.5   & 74.6 ${\color{darkred}\pm}$ 1.0 $^{\rm ^{\scalebox{0.99}{$\dagger$}}}$  &  37 & ... & 0.65 & 13, 14, 15  \\

\multicolumn{2}{l}{XTE J1650$-$500} &  {\color{blue}8} ${\color{darkred}\pm}$ {\color{blue}1.5}  &  {\color{blue}8} ${\color{darkred}\pm}$ {\color{blue}2}  & > 50 $^{\rm ^{\scalebox{0.99}{$\dagger$}}}$  &   7.63  & ... & 0.57 & 16, 17   \\ 

\multicolumn{2}{l}{XTE J1720$-$318} & {\color{blue}8} ${\color{darkred}\pm}$ {\color{blue}1.5}   &  {\color{blue}8} ${\color{darkred}\pm}$ {\color{blue}2} & ...  & ... & ... & 1.2 & 18, 19 \\ 
 
\multicolumn{2}{l}{XTE J1748$-$288} &  {\color{blue}8} ${\color{darkred}\pm}$ {\color{blue}1.5}  &  {\color{blue}8} ${\color{darkred}\pm}$ {\color{blue}2}   & ... &     ... & ... & 12 & 20 \\  
%\multicolumn{2}{l}{XTE J2012$+$381} &  {\color{darkred}8 $\pm$ 1.5}  &  {\color{darkred}8 $\pm$ 2}  & 40 &     ... & ... & 1.3 & 21, 22  \\ (21) \cite{Yao2005ApJ}, (22) \cite{White1998IAUC}

\multicolumn{2}{l}{XTE J1817$-$330} &  {\color{blue}8} ${\color{darkred}\pm}$ {\color{blue}1.5}  &  {\color{blue}8} ${\color{darkred}\pm}$ {\color{blue}2}  & ... &  ... & ... & 0.15 & 21  \\ 

\multicolumn{2}{l}{XTE J1908$+$094} & {\color{blue}8} ${\color{darkred}\pm}$ {\color{blue}1.5}   &  {\color{blue}8} ${\color{darkred}\pm}$ {\color{blue}2}   & 27.7 $\pm$ 3.4 $^{\rm ^{\scalebox{0.99}{$\ddagger$}}}$  & ... & ... & 2.3 & 22, 23 \\ 

\multicolumn{2}{l}{XTE J1752$-$223} &  {\color{blue}8} ${\color{darkred}\pm}$ {\color{blue}1.5}   &  {\color{blue}8} ${\color{darkred}\pm}$ {\color{blue}2}   & < 49 $^{\rm ^{\scalebox{0.99}{$\star$}}}$ &  < 6.8   & ... & 0.67 & 24  \\ 
\hline
\end{tabular}
\\[0.1cm]
\begin{tablenotes}\footnotesize \small
\item {\color{blue}a} \hspace{0.5cm} Binary separation, in lightseconds. 
\item {\color{blue}b} \hspace{0.5cm} \textbf{References} (1) \cite{Park2004ApJP}, (2) \cite{Jonker2004MNRAS}, (3) \cite{Orosz1998ApJ}, (4) \cite{Beer2002MNRAS}, (5) \cite{Orosz1997ApJ}, (6) \cite{2001MNRASGierlinski}, (7) \cite{Parker2016ApJ}, (8) \cite{Kong2000MNRAS}, (9) \cite{Zdziarski1998MNRAS}, (10) \cite{Gandhi2008MNRAS}, (11) \cite{Steiner2012ApJ}, (12) \cite{Blum2009cfdd}, (13) \cite{Orosz2011ApJ}, (14) \cite{Steiner2012ApJb}, (15) \cite{Gierli2003MNRAS}, (16) \cite{Homan2006MNRAS}, (17) \cite{Tomsick2004ApJ}, (18) \cite{Chaty2006A&A}, (19) \cite{Cadolle2004A&A}, (20) \cite{Revnivtsev2000MNRAS}, (21) \cite{Rykoff2007ApJ}, (22) \cite{miller2002broad}, (23) \cite{Zhang2015ApJ}, (24) \cite{miller2011accurate}
\item {\color{blue}c} \hspace{0.5cm} \cite{Shahbaz2003MNRAS} reported an alternative mass of 6.0 $\pm$ 0.4
\item $\dagger$ \hspace{0.5cm} obtained via ellipsoidal modulations
\item $\ddagger$ \hspace{0.5cm} obtained via reflection fitting 
\item $\star$ \hspace{0.5cm} obtained via proper motion of jet knots 

\end{tablenotes}
\end{table*}

\subsection{Source selection}\label{sec:datareduction}
We started our analyses while the sources are still in the soft state and followed their spectral and timing evolution towards the decay. In this sense, we have excluded hard-only outbursts. The complete list of the sources investigated in this study is given in \autoref{tab:bhtransients} alongside relevant observational properties. We have excluded 4U 1630$-$47 because of its erratic outburst behavior \citep{Tomsick05,Abe05}, contaminated background by the Galactic ridge and a dust scattering halo \citep{Kalemci18}.
For the sources without a mass measurement we have adopted a general value of 8 $\pm$ 1.5 M$_{\odot}$ (shown in blue in \autoref{tab:bhtransients}) based on work of \cite{ozel2010ApJ} and \cite{Kr2012ApJ}. 

\begin{figure}
\includegraphics[width=81mm]{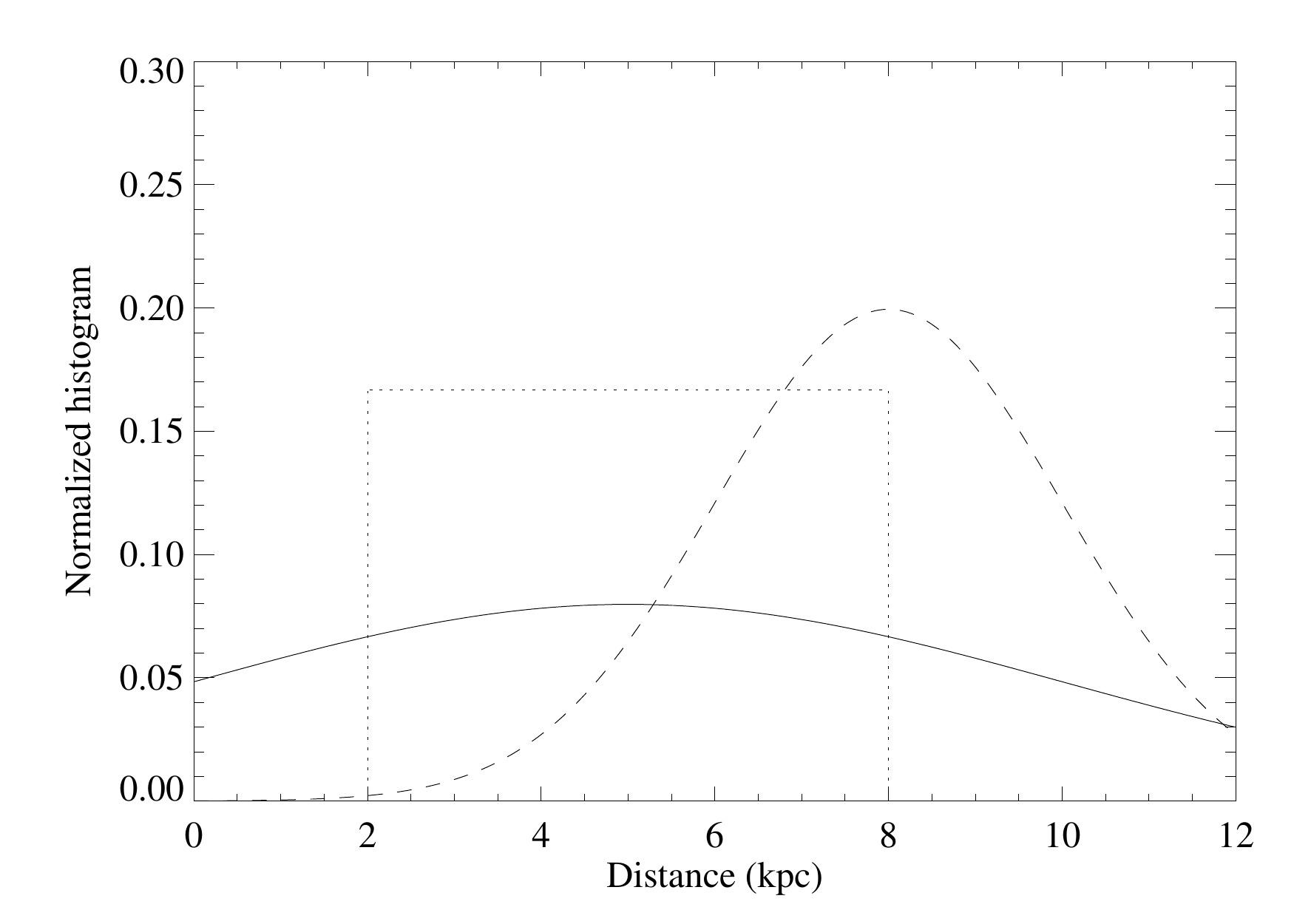}
\vspace{-0.30 cm}
\caption{\label{fig:distdist}
Distance distributions used in \protect\cite{dunn2010global}, \protect\cite{Tetarenko2016ApJS} and this  work are shown with solid, dotted and dashed lines, respectively.
}
\end{figure}

Source distance is one of the most important parameters in evaluating the Eddington fractions because it is either poorly determined (5\%-10\% error level, see \autoref{tab:bhtransients}), or not determined at all. Since the distance enters the luminosity calculation as a square, it is a major source of error. \cite{dunn2010global} and \cite{Tetarenko2016ApJS} used similar approaches to determine the effect of source distance by assuming a certain distribution of distances and drawing from these distributions many possible distances to obtain state transition luminosity distributions. While the methodology is similar, the distributions they assumed are different, \cite{dunn2010global} used 5 $\pm$ 5 kpc as distance distribution, while \cite{Tetarenko2016ApJS} uses a uniform distribution between 3 kpc and 8kpc. Figure~\ref{fig:distdist} shows the differences between these distributions.

\begin{figure}
\includegraphics[width=90mm]{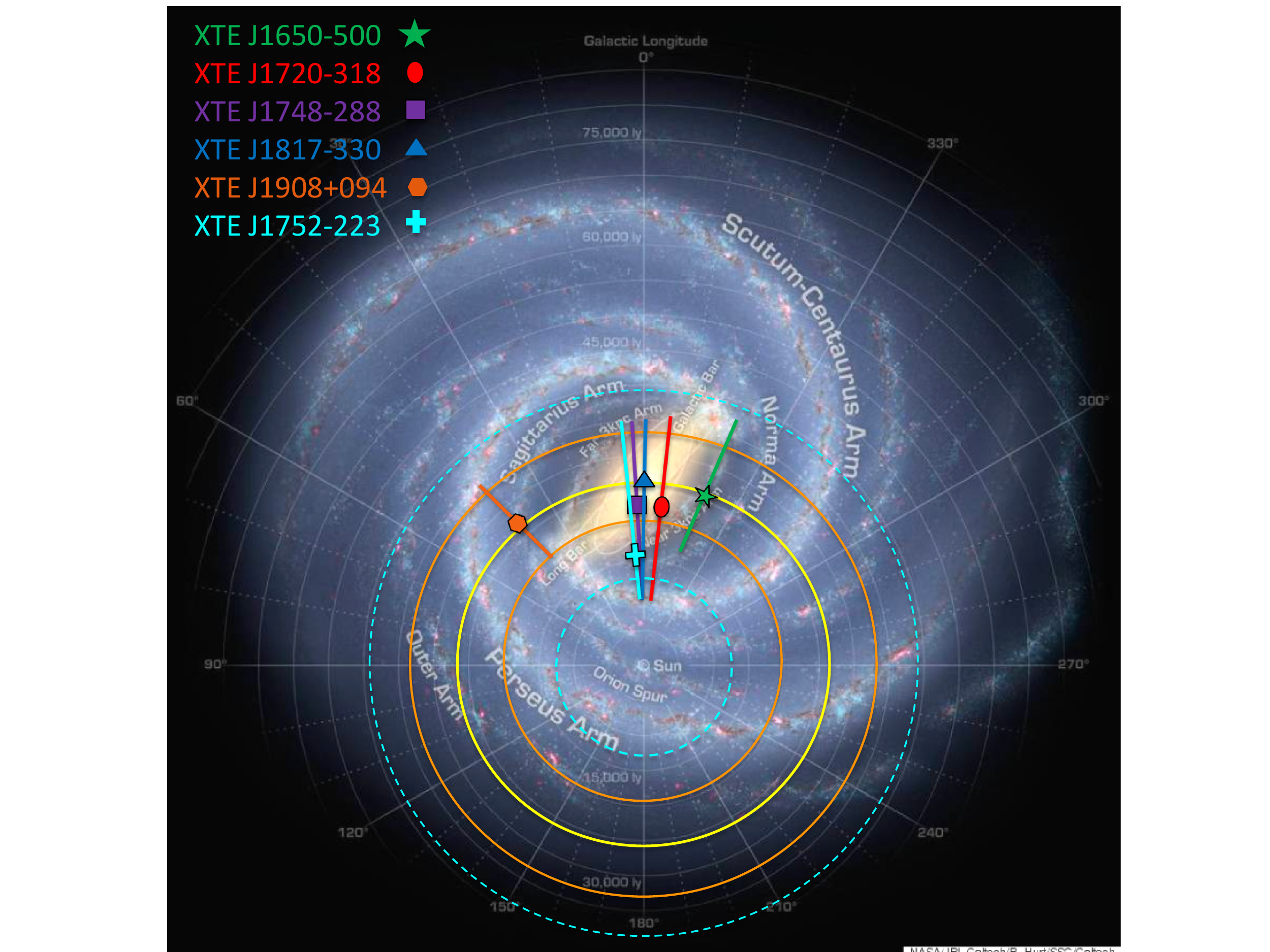}
\vspace{-0.50 cm}
\caption{\label{fig:distest3}
The Galactic distribution of black holes with unknown distance used in this study. The yellow circle displays the 8 kpc radius. Orange circles indicate 1$\sigma$ range used in this work (8 $\pm$ 2 kpc) whereas blue dashed lines indicate 2$\sigma$ range. The symbols do not represent the maximum likelihood value but placed just for presentation purposes. Error bars are arranged from arm to arm or arm toward the Galactic Bulge.  (Milky Way image: NASA/JPL-Caltech, ESO, J. Hurt.)
}
\end{figure}

\cite{CorralSantana16} cataloged all known black holes and candidates at the time of publication and analyzed the distribution of sources in the Galaxy. In Fig.~\ref{fig:distest3} we took their Fig.~2 that shows the distribution of sources in the Galaxy and overlaid our distance estimate of 8 $\pm$ 2 kpc as well as sources without distance by using Galactic coordinates. According to \cite{CorralSantana16}, the sample of GBHTs is complete out to $r\sim$ 4 kpc. 

Although there are most likely closer quiescent transients that were not detected in outburst, we choose to take this as an indication that the sources with unknown mass and distance measurements are likely to be at distances greater than 4 kpc. Moreover, \cite{CorralSantana16} also pointed out the fact that all distance determined GBHTs are in the Galactic bulge or in the spiral arms. We note that most of the unknown distance sources lie in the Galactic bulge direction. It is likely that a large fraction of those sources are in the Galactic bulge. As seen in Fig.~\ref{fig:distest3}, a distance estimate of 8 $\pm$ 2 kpc engulfs almost the entire bulge within 1$\sigma$.

%% /*******************************************************************
%% ** Spectral state definitions                                     **
%% *******************************************************************/

\subsection{Spectral state definitions}\label{sec:states}

\begin{figure}
\includegraphics[width=85mm]{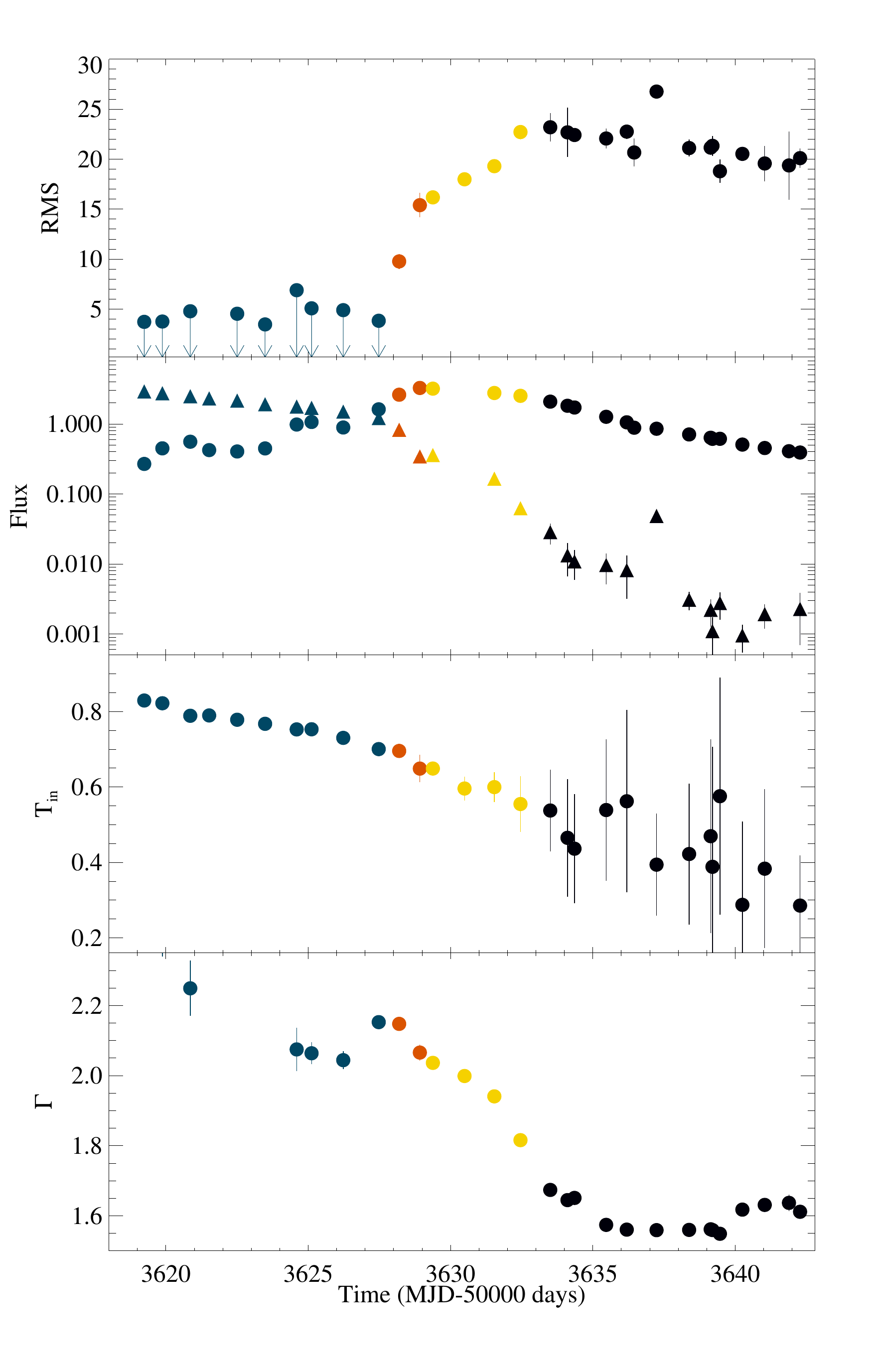}
\vspace{-0.40 cm}
\caption{\label{fig:J1655}
The X-ray spectral and temporal parameters of GRO J1655$-$40 in 2005 outburst decay as a function of time. (a) rms variability, (b) DBB (shown with triangles) and power-law flux (shown with circles)(in 10$^{-9}$ ergs s$^{-1}$ cm$^{-2}$), (c) inner disk temperature (in keV) and (d) X-ray photon index. In the parameter evolution, different states are colored according to classification reported in \autoref{tab:kalemci13}. The observations before timing transition are shown with blue in figures. This transition is also the reference date in \autoref{tab:kalemci13} where other transitions are stated. The observations after the timing transition but before the index transition are presented with orange. The observations before compact jet transition and after the index transition are shown with yellow. Lastly, all observations after the compact jet transition are presented in black.}
\end{figure}

\begin{figure}
\includegraphics[width=85mm]{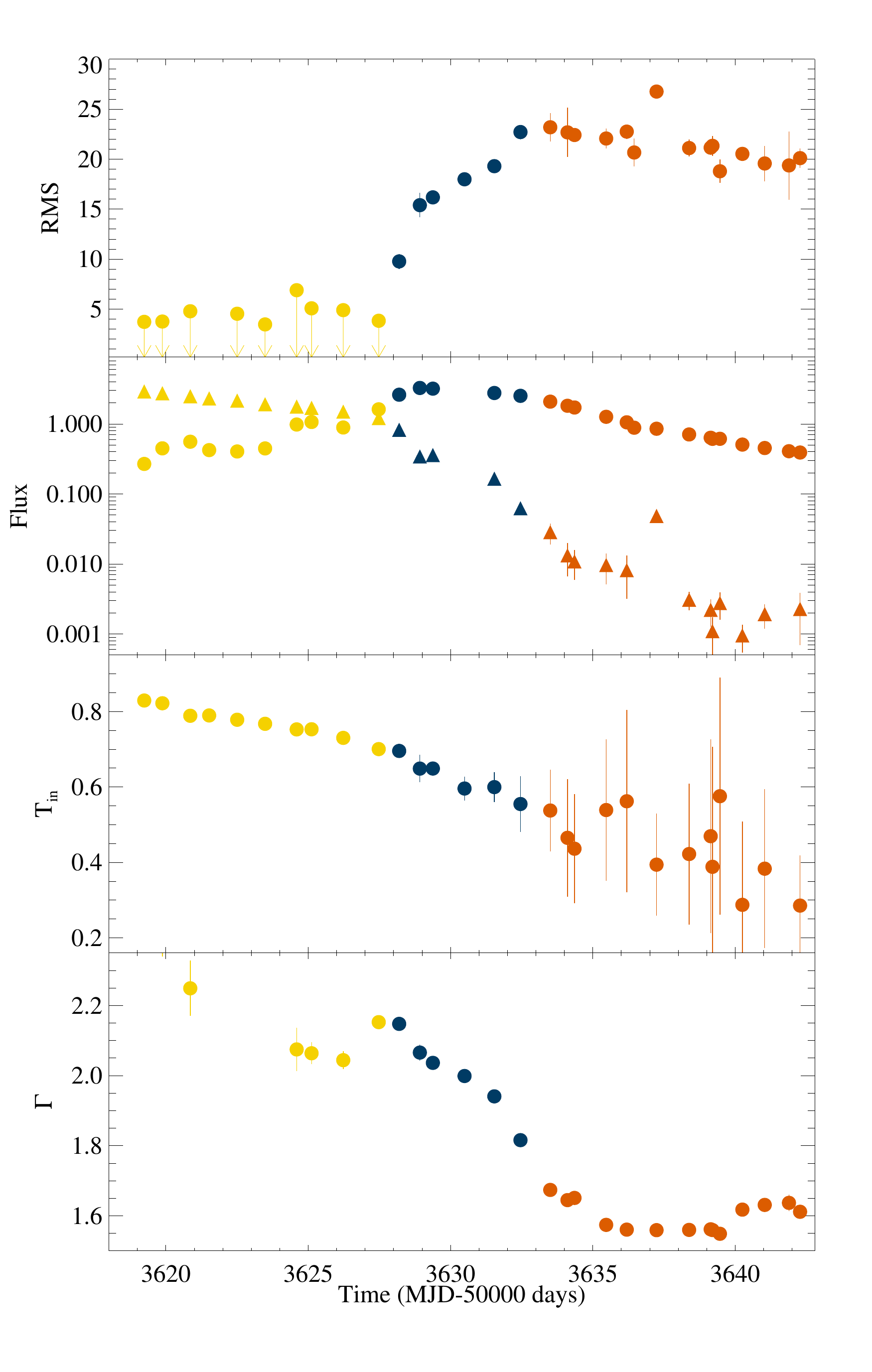}
\vspace{-0.5 cm}
\caption{\label{fig:J1655K}
The X-ray spectral and temporal parameters of GRO J1655$-$40 in 2005 outburst decay as a function of time. (a) rms variability, (b) DBB (shown with triangles) and power-law flux (shown with circles)(in 10$^{-9}$ erg s$^{-1}$ cm$^{-2}$), (c) inner disk temperature (in keV) and (d) X-ray photon index. In the parameter evolution, different states are colored according to Belloni classification (\autoref{tab:belloni}) where soft states are presented with yellow colour whereas the hard intermediate (HIMS) and hard states are displayed with blue and orange colour respectively.}
\end{figure}

\begin{figure}
\includegraphics[width=85mm]{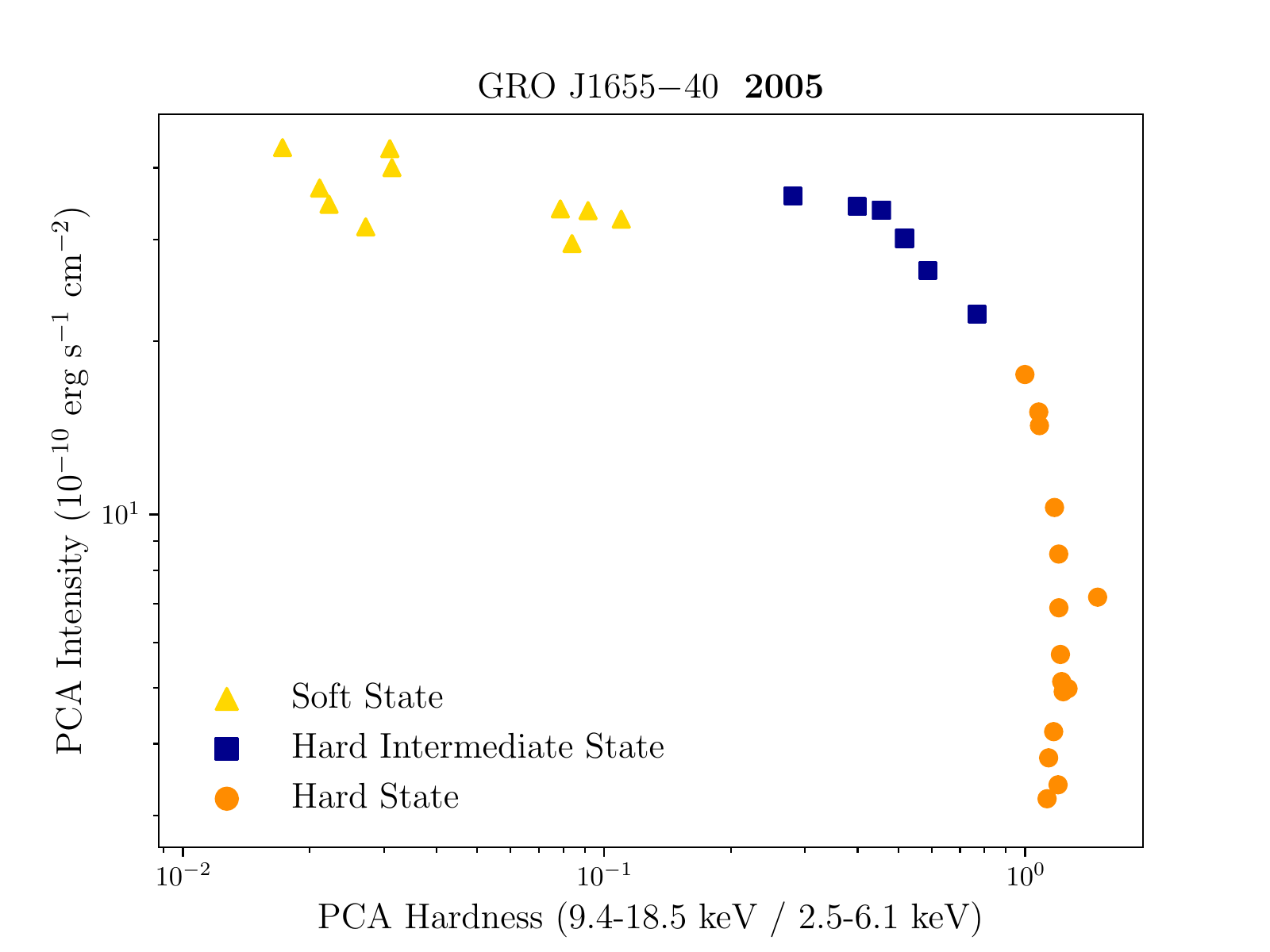}
\vspace{-0.50 cm}
\caption{\label{fig:GRO2005_HID}
Hardness intensity diagram of GRO J1655$-$40 in 2005 outburst decay.  Different states are colored according to Belloni classification (\autoref{tab:belloni}) where soft states are presented with yellow triangle whereas the hard intermediate (HIMS) and hard states are displayed with a blue square and orange circle respectively.}
\end{figure}

The state transition luminosities and their corresponding transition times are determined and given in \autoref{tab:kalemci13} and \autoref{tab:belloni} using the classifications discussed in \S\ref{stl}. 2005 outburst of GRO J1655$-$40 has chosen to be a reference case for determining the state transitions due to precise mass and distance measurements as well as good spectral and temporal coverage. 

We have examined 30 observations in total for the 2005 outburst of GRO J1655$-$40 between MJD 53619$-$53644 until the source decayed into quiescence. As it can be seen from \autoref{fig:J1655} and \autoref{fig:J1655K}, the DBB flux (b) dropped below the power-law flux on MJD 53626 (transition to HIMS). This also corresponds to the time where the rms in 2$-$30 keV started to increase (timing transition) and continued up to 34\%. Power spectra were best fitted with 2 broad and a single narrow Lorentzian (a type-C QPO around 10 Hz). After MJD 53639, the QPO disappeared and two broad Lorentzians were enough to model the spectra. The photon index (d) also dropped from $\sim$2.1 to $\sim$1.7 in this period (index transition) and flattened on MJD 53634 (compact jet transition and the hard state). The source also showed softening after MJD 53641 which is not taken into account as an additional state transition in this work because not only softening at the luminosity levels we are interested in is not universal in GBHT decays, but also very difficult to distinguish from artificial softening due to Galactic Ridge contribution. For example, H1743$-$322 is a GBHT that the effect of Galactic ridge emission is studied due to the proximity of the source to the Galactic Plane. \cite{Kalemci2006ApJK} have reported a constant unabsorbed flux of 1.08 $\times$ 10$^{-10}$ ergs s$^{-1}$ cm$^{-2}$ from the ridge emission during 2003 outburst decay which contributes 10\% of the total flux during the transition to the hard state.

As one can see in Figures~\ref{fig:J1655} and \ref{fig:J1655K}, in the majority of the cases, there is an overlap between the timing transition and the transition to HIMS as well as between the compact jet transition and the transition to the hard state. Therefore, we have decided to use the \cite{Belloni10_jp} classification and just added the index transition from \cite{kalemci2013complete} as an additional state transition to our analysis. The hardness-intensity diagram of this outburst with respect to \cite{Belloni10_jp} classification has also been given in \autoref{fig:GRO2005_HID}.  We represented the HR as the ratio of PCA flux in the energy bands 9.4 -- 18.5 keV and 2.5 -- 6.1 keV and the intensity as the PCA flux in the band 2.5 -- 18.5 keV \citep{Corbel2013}.

For the rest of outbursts, we have followed a similar procedure for identifying the state transitions. If there are published data regarding state transitions, we compared them to our results. In general, the state transition dates found in the literature are similar to our findings.

%\subsection{Mass and distance measurements}\label{sec:massdist}

\subsection{Bolometric correction}\label{sec:bolcor}

A bolometric correction is a necessary step for the accurate determination of the Eddington scaled luminosities. We have applied an X-ray bolometric correction to the data as follows: If the disk is detected, then the DBB luminosity is calculated using the range of 0.01 -- 200 keV, and the power-law luminosity is calculated from the disk temperature $T_{\rm in}$ to 200 keV from the present model parameters. At low luminosities in the hard state, the disk is often not detected. We choose a range from 0.5 -- 200 keV as the range to calculate power-law luminosities. When HEXTE is present we also include the effect of the high energy cut-off, whereas for the PCA-only cases no cut-off is included in the bolometric correction.

We have also tried bolometric corrections adopted in previous studies and re-calculated our results accordingly. \cite{maccarone2003x} assumed a spectrum of $\frac{\rm dN}{\rm dE}$ $\sim$ E$^{-1.8}$ exp$^{-E/200 ~ \rm{keV}}$ from 0.5 keV to 10 MeV for majority of sources in the hard state regardless of the individual spectral fits. For our reference source, all the key parameters agreed within $\approx$ 10\%.
\cite{dunn2010global} implemented a correction for the disk and power-law from 0.001 and 1 keV up to 100 keV, respectively. \cite{Tetarenko2016ApJS} have computed the 2 -- 50 keV flux for each observation using a Monte Carlo algorithm and converted to bolometric flux in 0.001 -- 1000 keV band by multiplying disk and power-law component by a derived bolometric correction from the XSPEC models. Although PL luminosities we calculated agreed within 15\% on average for various states with those of \cite{Tetarenko2016ApJS} and \cite{dunn2010global}, the DBB luminosities of \cite{Tetarenko2016ApJS} are up to one order of magnitude lower than those of \cite{dunn2010global} and our study.

\begin{figure}
\includegraphics[width=85mm]{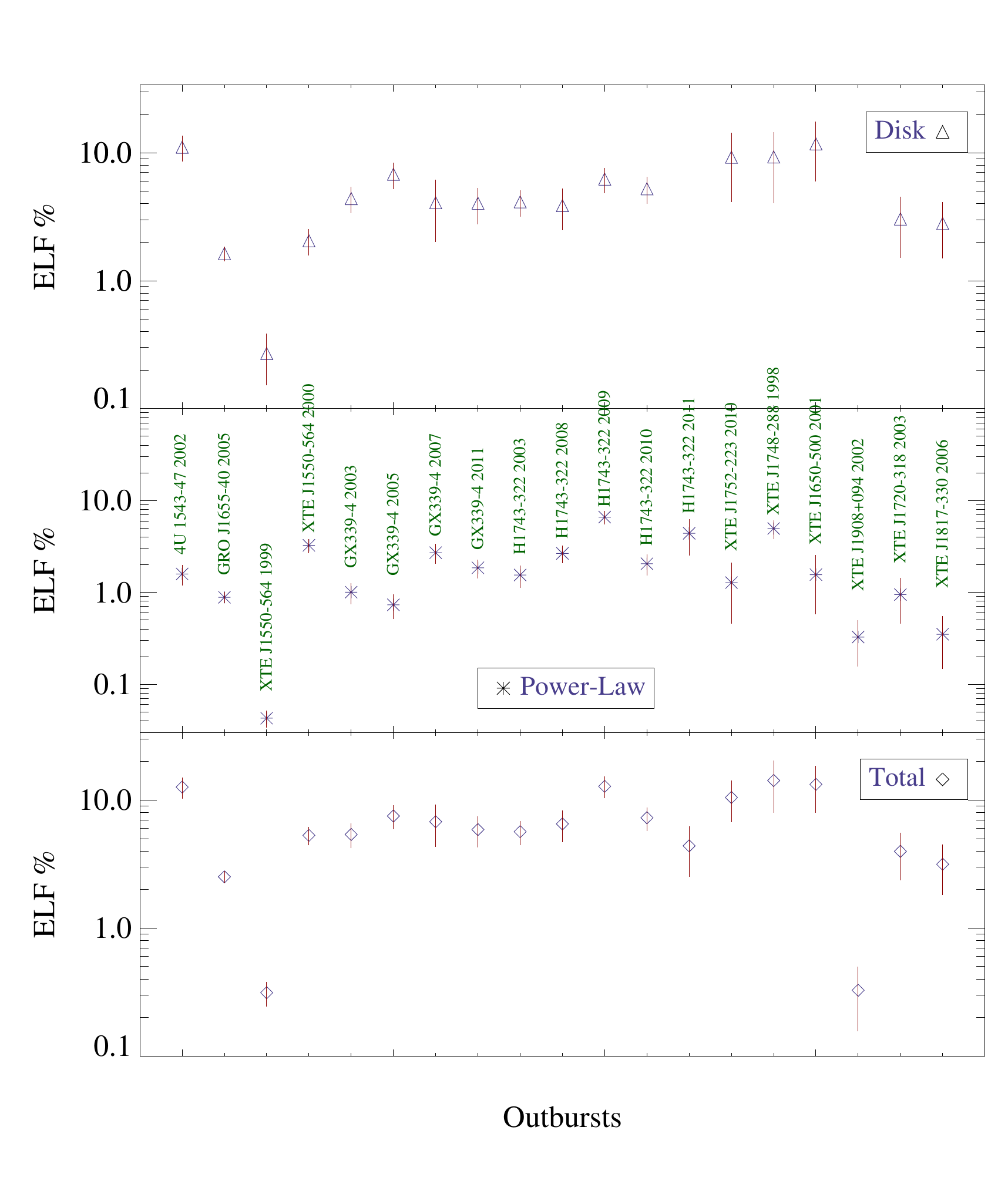}
\vspace{-1.0 cm}
\caption{\label{fig:trans_hims}
ELF distribution of GBHTs during transition to HIMS. \textit{Top:} DBB, \textit{middle:} PL, \textit{bottom:} Total.
}
\end{figure}

\begin{figure}
\includegraphics[width=85mm]{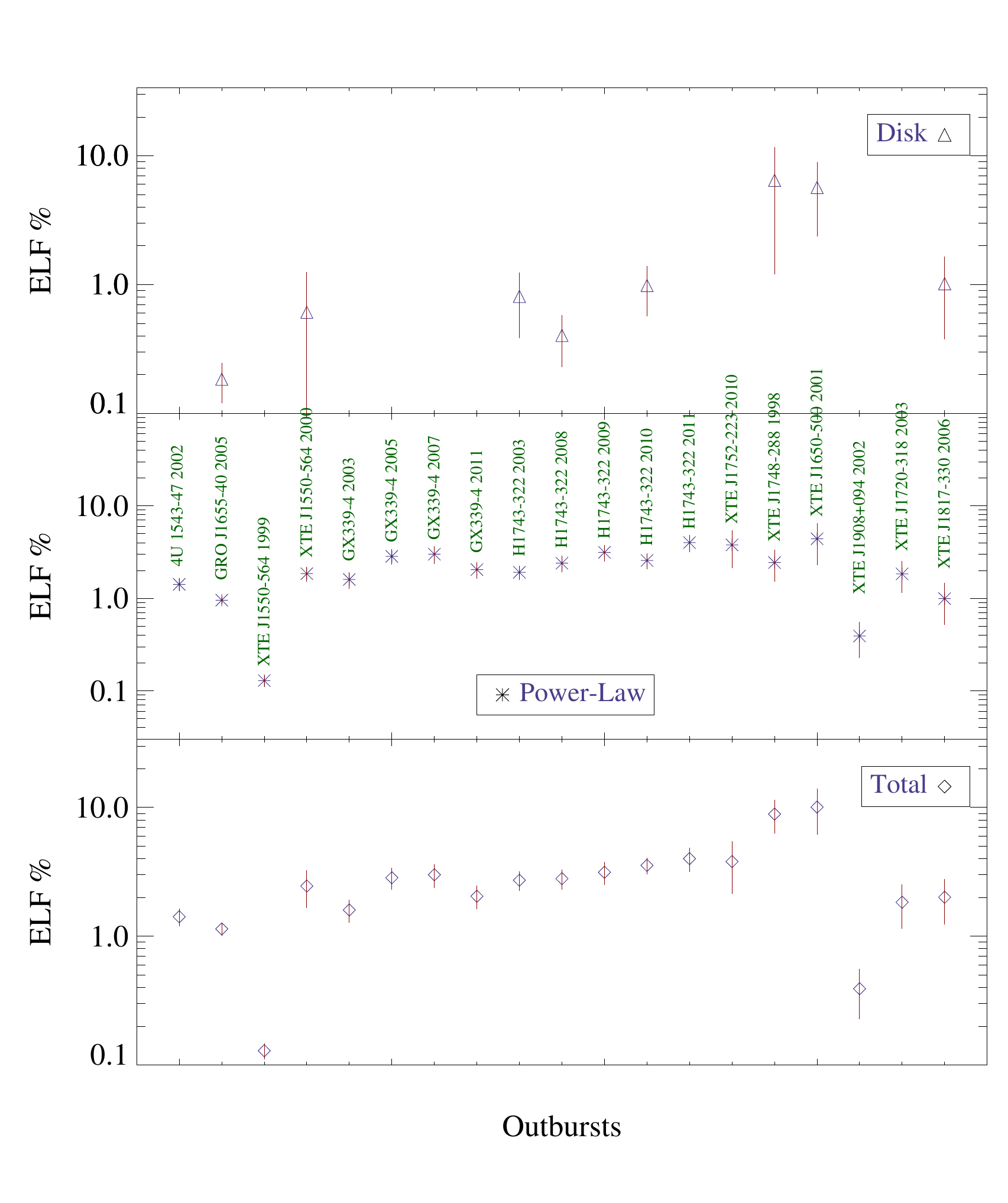}
\vspace{-1.0 cm}
\caption{\label{fig:trans_hs}
ELF distribution of GBHTs during transition to the hard state. \textit{Top:} DBB , \textit{middle:} PL , \textit{bottom:} Total .
}
\end{figure}

%% /*******************************************************************
%% ** Results                                                        **
%% *******************************************************************/

\section{Results}\label{sec:results}

We have calculated the state transition luminosities for 11 GBHT in 19 different outbursts in terms of Eddington luminosity fractions (ELF), for state transition definitions according to both \cite{Belloni10_jp} and \cite{kalemci2013complete}. We then listed bolometrically corrected luminosities and their occurrence times in \autoref{tab:kalemci13} and \autoref{tab:belloni}. We have also separated disk and power-law luminosity fractions. 
\vspace{1cm}
\begin{table*}
\centering
\caption[]{Transition times and Eddington luminosity fractions (PL+DBB) based on \cite{kalemci2013complete}\label{tab:kalemci13}.}
\scalebox{0.95}{
\begin{tabular}{l|cc|cc|cc}
\hline \hline
 & \multicolumn{2}{c}{Timing Transition (TT)} & \multicolumn{2}{c}{Index Transition (IT)} & \multicolumn{2}{c}{Compact Jet Transition (CJT)} \Tstrut\Bstrut \\
Source, Year  &  Date & ELF & Lag$ ^{\rm ^{\scalebox{0.96}{{\color{blue}a}}}}$ & ELF & Lag & ELF \\
 & (MJD) & (\%) & (days) & (\%) & (days) & (\%) \\
\hline \Tstrut 
4U 1543-47, 2002 & 52473.7 ${\color{darkred}\pm}$ 0.4 & 12.65 ${\color{darkred}\pm}$ 2.95 &  6.5 ${\color{darkred}\pm}$ 0.4 & 8.25 ${\color{darkred}\pm}$ 2.69 &  9.7 ${\color{darkred}\pm}$ 0.5 & 1.42 ${\color{darkred}\pm}$ 0.28 \\
GRO J1655-40, 2005 & 53627.8 ${\color{darkred}\pm}$ 0.3 & 2.52 ${\color{darkred}\pm}$ 0.33 &  1.3 ${\color{darkred}\pm}$ 0.2 & 2.01 ${\color{darkred}\pm}$ 0.27 & 3.9 ${\color{darkred}\pm}$ 0.5  & 1.14 ${\color{darkred}\pm}$ 0.16 \\
XTE J1550-564, 1999 & 51306.2 ${\color{darkred}\pm}$ 1.1 & 0.19 ${\color{darkred}\pm}$ 0.03 &  $-$ & $-$ &  2.3 ${\color{darkred}\pm}$ 1.2 & 0.13 ${\color{darkred}\pm}$ 0.02 \\
XTE J1550-564, 2000 & 51674.1 ${\color{darkred}\pm}$ 0.6 & 6.15 ${\color{darkred}\pm}$ 1.43 &  7.2 ${\color{darkred}\pm}$ 0.9 & 2.53 ${\color{darkred}\pm}$ 0.54 & 1.7 ${\color{darkred}\pm}$ 0.7 & 2.45 ${\color{darkred}\pm}$ 1.00\\
GX339-4, 2003 & 52717.8 ${\color{darkred}\pm}$ 0.3 & 5.39 ${\color{darkred}\pm}$ 1.47 &  7.9 ${\color{darkred}\pm}$ 1.5 & 3.85 ${\color{darkred}\pm}$ 0.60 & 22.0 ${\color{darkred}\pm}$ 0.2 & 1.68 ${\color{darkred}\pm}$ 0.44 \\
GX339-4, 2005 & 53461.3 ${\color{darkred}\pm}$ 1.8 & 7.81 ${\color{darkred}\pm}$ 2.48 &  6.7 ${\color{darkred}\pm}$ 1.2 & 5.10 ${\color{darkred}\pm}$ 0.81 & 14.7 ${\color{darkred}\pm}$ 1.7 & 2.85 ${\color{darkred}\pm}$ 0.67 \\
GX339-4, 2007 & 54228.0 ${\color{darkred}\pm}$ 0.4 & 6.83 ${\color{darkred}\pm}$ 1.81 &  5.7 ${\color{darkred}\pm}$ 3.4 & 6.32 ${\color{darkred}\pm}$ 1.61 & 1.3 ${\color{darkred}\pm}$ 0.6 & 3.43 ${\color{darkred}\pm}$ 0.88\\
GX339-4, 2011 & 55594.2 ${\color{darkred}\pm}$ 0.7 & 5.88 ${\color{darkred}\pm}$ 2.00 &  3.8 ${\color{darkred}\pm}$ 0.7 & 4.83 ${\color{darkred}\pm}$ 2.10 & 11.7 ${\color{darkred}\pm}$ 1.0 & 2.04 ${\color{darkred}\pm}$ 0.52\\
H1743-322, 2003 & 52930.4 ${\color{darkred}\pm}$ 0.5 & 5.67 ${\color{darkred}\pm}$ 1.53 &  4.5 ${\color{darkred}\pm}$ 1.0 & 4.17 ${\color{darkred}\pm}$ 1.09 & 10.2 ${\color{darkred}\pm}$ 1.5 & 2.72 ${\color{darkred}\pm}$ 0.59 \\
H1743-322, 2008 & 54488.3 ${\color{darkred}\pm}$ 0.9 & 6.52 ${\color{darkred}\pm}$ 2.27 &  9.0 ${\color{darkred}\pm}$ 0.5 & 3.52 ${\color{darkred}\pm}$ 1.03 & 11.5 ${\color{darkred}\pm}$ 0.9 & 2.80 ${\color{darkred}\pm}$ 0.61 \\
H1743-322, 2009 & 55014.7 ${\color{darkred}\pm}$ 1.6 & 4.21 ${\color{darkred}\pm}$ 0.88 &  9.5 ${\color{darkred}\pm}$ 1.0 & 2.62 ${\color{darkred}\pm}$ 0.62 & 12.4 ${\color{darkred}\pm}$ 1.8 & 1.98 ${\color{darkred}\pm}$ 0.43 \\
H1743-322, 2010 & 55449.5 ${\color{darkred}\pm}$ 0.8 & 5.41 ${\color{darkred}\pm}$ 1.40 &  1.3 ${\color{darkred}\pm}$ 0.4 & 5.03 ${\color{darkred}\pm}$ 1.30 & 6.6 ${\color{darkred}\pm}$ 5.1  & 3.55 ${\color{darkred}\pm}$ 0.64 \\
H1743-322, 2011 & 55678.7 ${\color{darkred}\pm}$ 0.6 & 8.45 ${\color{darkred}\pm}$ 2.06 &  7.4 ${\color{darkred}\pm}$ 1.4 & 6.57 ${\color{darkred}\pm}$ 1.80 & 12.1 ${\color{darkred}\pm}$ 0.7 & 4.01 ${\color{darkred}\pm}$ 1.04 \\
XTE J1752-223, 2010 & 55282.1 ${\color{darkred}\pm}$ 2.1 & 10.49 ${\color{darkred}\pm}$ 4.70 &  9.1 ${\color{darkred}\pm}$ 2.1 & 5.71 ${\color{darkred}\pm}$ 1.95 & 13.2 ${\color{darkred}\pm}$ 1.9 & 3.80 ${\color{darkred}\pm}$ 2.08 \\
XTE J1748-288, 1998 & 51009.7 ${\color{darkred}\pm}$ 2.4 & 14.24 ${\color{darkred}\pm}$ 7.79 &  $-$ &  $-$ & 8.6 ${\color{darkred}\pm}$ 6.1 & 8.89 ${\color{darkred}\pm}$ 3.27 \\

XTE J1650-500, 2001 & 52228.6 ${\color{darkred}\pm}$ 0.4 & 14.98 ${\color{darkred}\pm}$ 7.57 &  4.0 ${\color{darkred}\pm}$ 0.6 & 11.82 ${\color{darkred}\pm}$ 6.14 & 5.3 ${\color{darkred}\pm}$ 0.6 & 10.01 ${\color{darkred}\pm}$ 4.93 \\

XTE J1908+94, 2002 & 52427.5 ${\color{darkred}\pm}$ 2.0 & $-$ & 0.0 ${\color{darkred}\pm}$ 0.0 & 0.32 ${\color{darkred}\pm}$ 0.21  & 4.6 ${\color{darkred}\pm}$ 2.6 & 0.39 ${\color{darkred}\pm}$ 0.19 \\

XTE J1720-318, 2003 & 52726.6 ${\color{darkred}\pm}$ 2.8 & 2.88 ${\color{darkred}\pm}$ 1.09 & -10.3 ${\color{darkred}\pm}$ 1.3 & 4.79 ${\color{darkred}\pm}$ 2.52 & 22.0 ${\color{darkred}\pm}$ 1.0 & 0.98 ${\color{darkred}\pm}$ 0.50 \\

XTE J1817-330,2006 & 53885.3 ${\color{darkred}\pm}$ 1.8 & 3.15 ${\color{darkred}\pm}$ 1.67 & 2.9 ${\color{darkred}\pm}$ 1.1 & 2.81 ${\color{darkred}\pm}$ 1.40 & 7.3 ${\color{darkred}\pm}$ 1.9 & 2.01 ${\color{darkred}\pm}$ 0.97\\

\end{tabular}
}
\begin{tablenotes}\footnotesize \small \vspace{+1mm}
\item {\color{blue}a} \hspace{0.5cm} All lags are with respect to the timing transition.
\end{tablenotes}

\end{table*}

\begin{table*}
\centering
\caption[]{Transition times and Eddington luminosity fractions (PL+DBB) based on \cite{Belloni10_jp}. \label{tab:belloni}}
\scalebox{0.95}{
\begin{tabular}{l|cc|cc|cc}
\hline \hline
 & \multicolumn{2}{c}{Soft Intermediate (SIMS)} & \multicolumn{2}{c}{Hard Intermediate (HIMS)} & \multicolumn{2}{c}{Hard (HS)} \Tstrut\Bstrut \\
Source, Year  &  Lag$ ^{\rm ^{\scalebox{0.96}{{\color{blue}a}}}}$ & ELF &  Date & ELF & Lag & ELF \\
 & (days) & (\%) &  (MJD) & (\%) & (days) & (\%) \\
\hline \Tstrut 
4U 1543-47, 2002 & $-$ & $-$ &  52473.7 ${\color{darkred}\pm}$ 0.4 & 12.65 ${\color{darkred}\pm}$ 2.06 &  9.7 ${\color{darkred}\pm}$ 0.5 & 1.42 ${\color{darkred}\pm}$ 0.19 \\
GRO J1655-40, 2005 & $-$ & $-$ &  53627.8 ${\color{darkred}\pm}$ 0.3 & 2.51 ${\color{darkred}\pm}$ 0.33 & 5.2 ${\color{darkred}\pm}$ 0.5  & 1.14 ${\color{darkred}\pm}$ 0.16 \\
XTE J1550-564, 1999 & $-$ & $-$ &  51304.3 ${\color{darkred}\pm}$ 0.8 & 0.31 ${\color{darkred}\pm}$ 0.08 &  4.2 ${\color{darkred}\pm}$ 1.2 & 0.12 ${\color{darkred}\pm}$ 0.02 \\
XTE J1550-564, 2000 & -9.7 ${\color{darkred}\pm}$ 0.5 & 3.95 ${\color{darkred}\pm}$ 1.81 &  51675.1 ${\color{darkred}\pm}$ 0.4 & 5.30 ${\color{darkred}\pm}$ 1.09 & 7.9 ${\color{darkred}\pm}$ 0.7 & 2.45 ${\color{darkred}\pm}$ 1.00 \\
GX339-4, 2003 & $-$ & $-$ &  52717.8 ${\color{darkred}\pm}$ 0.3 & 5.39 ${\color{darkred}\pm}$ 1.47 & 23.1 ${\color{darkred}\pm}$ 0.8 & 1.60 ${\color{darkred}\pm}$ 0.40 \\
GX339-4, 2005 & $-$ & $-$ &  53457.6 ${\color{darkred}\pm}$ 1.8 & 7.52 ${\color{darkred}\pm}$ 1.98 & 18.4 ${\color{darkred}\pm}$ 1.8 & 2.85 ${\color{darkred}\pm}$ 0.67 \\
GX339-4, 2007 & -9.0 ${\color{darkred}\pm}$ 0.5 & 3.86 ${\color{darkred}\pm}$ 2.09 &  54234.2 ${\color{darkred}\pm}$ 0.6 & 6.78 ${\color{darkred}\pm}$ 3.07 & 3.4 ${\color{darkred}\pm}$ 0.1 & 3.00 ${\color{darkred}\pm}$ 0.79\\
GX339-4, 2011 & -9.0 ${\color{darkred}\pm}$ 0.7 & 6.47 ${\color{darkred}\pm}$ 1.75 &  55594.2 ${\color{darkred}\pm}$ 0.7 & 5.88 ${\color{darkred}\pm}$ 2.00 & 11.7 ${\color{darkred}\pm}$ 1.0 & 2.04 ${\color{darkred}\pm}$ 0.52\\
H1743-322, 2003 & $-$ & $-$ &  52930.4 ${\color{darkred}\pm}$ 0.4 & 5.67 ${\color{darkred}\pm}$ 1.53 & 10.2 ${\color{darkred}\pm}$ 1.5 & 2.72 ${\color{darkred}\pm}$ 0.59  \\
H1743-322, 2008 & $-$ & $-$ &  54488.3 ${\color{darkred}\pm}$ 0.9 & 6.52 ${\color{darkred}\pm}$ 2.27 & 11.5 ${\color{darkred}\pm}$ 1.0  & 2.80 ${\color{darkred}\pm}$ 0.61 \\
H1743-322, 2009 & -4.8 ${\color{darkred}\pm}$ 0.5 & 6.91 ${\color{darkred}\pm}$ 3.67 &  54991.1 ${\color{darkred}\pm}$ 0.6 & 7.01 ${\color{darkred}\pm}$ 3,40 & 26.8 ${\color{darkred}\pm}$ 1.5 & 3.14 ${\color{darkred}\pm}$ 0.79 \\
H1743-322, 2010 & $-$ & $-$ &  55441.6 ${\color{darkred}\pm}$ 1.3 & 7.29 ${\color{darkred}\pm}$ 1.89 & 15.8 ${\color{darkred}\pm}$ 5.1 & 3.55 ${\color{darkred}\pm}$ 0.64 \\
H1743-322, 2011 & $-$ & $-$ &  55670.7 ${\color{darkred}\pm}$ 0.5 & 4.39 ${\color{darkred}\pm}$ 2.33 & 20.1 ${\color{darkred}\pm}$ 0.7 & 4.01 ${\color{darkred}\pm}$ 1.04 \\
XTE J1752-223, 2010 & $-$ & $-$ &  55281.8 ${\color{darkred}\pm}$ 2.1 & 10.49 ${\color{darkred}\pm}$ 4.70 & 13.2 ${\color{darkred}\pm}$ 1.9 & 3.80 ${\color{darkred}\pm}$ 2.08 \\
XTE J1748-288, 1998 & $-$ & $-$ &  51009.7 ${\color{darkred}\pm}$ 2.4 & 14.23 ${\color{darkred}\pm}$ 7.79 & 8.6 ${\color{darkred}\pm}$ 6.1 & 8.89 ${\color{darkred}\pm}$ 3.27 \\

XTE J1650-500, 2001 & $-$ & $-$ &  52231.5 ${\color{darkred}\pm}$ 0.5 & 13.29 ${\color{darkred}\pm}$ 6.64 & 2.4 ${\color{darkred}\pm}$ 0.6 & 10.01 ${\color{darkred}\pm}$ 4.93 \\

XTE J1908+94, 2002 & $-$ & $-$ &  52427.5 ${\color{darkred}\pm}$ 2.0 & 0.32 ${\color{darkred}\pm}$ 0.21 & 4.6 ${\color{darkred}\pm}$ 2.6 & 0.39 ${\color{darkred}\pm}$ 0.20 \\

XTE J1720-318, 2003 & $-$ & $-$ & 52728.2 ${\color{darkred}\pm}$ 1.5 & 3.98 ${\color{darkred}\pm}$ 2.01 & 9.4 ${\color{darkred}\pm}$ 2.1 & 1.83 ${\color{darkred}\pm}$ 0.87 \\

XTE J1817-330, 2006 & $-$ & $-$ & 53885.3 ${\color{darkred}\pm}$ 1.8 & 3.16 ${\color{darkred}\pm}$ 1.66 & 7.3 ${\color{darkred}\pm}$ 1.2 & 2.01 ${\color{darkred}\pm}$ 0.97 \\

\end{tabular}
}
\begin{tablenotes}\footnotesize \small \vspace{+1mm}
\item {\color{blue}a} \hspace{0.5cm} All lags are with respect to the HIMS transition.
\end{tablenotes}

\end{table*}

\vspace{1cm}
\begin{table*}
\centering
\caption[]{Average state transition luminosity of GBHTs in terms of Eddington luminosity fractions.}\label{tab:wmean}
\scalebox{0.95}{
\begin{tabular}{l|cc|cc|cc}
\hline \hline
 & \multicolumn{2}{c}{Weighted mean 1$ ^{\rm ^{\scalebox{0.96}{{\color{blue}a}}}}$} & \multicolumn{2}{c}{Weighted mean 2$ ^{\rm ^{\scalebox{0.96}{{\color{blue}b}}}}$} & \multicolumn{2}{c}{Histogram$ ^{\rm ^{\scalebox{0.96}{{\color{blue}c}}}}$} \Tstrut\Bstrut \\
Transition type$^{^{{\color{blue}\ast}}}$  &  DBB & PLF & DBB & PLF & DBB & PLF \\
 & (\%) & (\%) & (\%) & (\%) & (\%) & (\%) \\
\hline \Tstrut  \\
SIMS & 2.79 ${\color{darkred}\pm}$ 0.50 & 1.88 ${\color{darkred}\pm}$ 0.35 &  2.94 ${\color{darkred}\pm}$ 0.53 & 2.79 ${\color{darkred}\pm}$ 0.53 &  -- & -- \\\\

SIMS $-$ o & 2.79 ${\color{darkred}\pm}$ 0.50 & 1.88 ${\color{darkred}\pm}$ 0.35 &  2.94 ${\color{darkred}\pm}$ 0.53 & 2.79 ${\color{darkred}\pm}$ 0.52 &  -- & -- \\\\

HIMS  & 0.78 ${\color{darkred}\pm}$ 0.12 & 0.05 ${\color{darkred}\pm}$ 0.01 &  4.38 ${\color{darkred}\pm}$ 0.51 & 2.07 ${\color{darkred}\pm}$ 0.22 &  -1.50 ${\color{darkred}\pm}$ 0.32 & -1.87 ${\color{darkred}\pm}$ 0.37
 \\\\
 
HIMS $-$ o & 2.03 ${\color{darkred}\pm}$ 0.23 & 0.79 ${\color{darkred}\pm}$ 0.09 &  4.51 ${\color{darkred}\pm}$ 0.54 & 2.33 ${\color{darkred}\pm}$ 0.25 & -1.49 ${\color{darkred}\pm}$ 0.30 & -1.87 ${\color{darkred}\pm}$ 0.40 \\\\

IT  & 0.33 ${\color{darkred}\pm}$ 0.05 & 0.21 ${\color{darkred}\pm}$ 0.03 &  2.05 ${\color{darkred}\pm}$ 0.30 & 1.67 ${\color{darkred}\pm}$ 0.15 & -1.83 ${\color{darkred}\pm}$ 0.48 & -1.70 ${\color{darkred}\pm}$ 0.21 \\\\

IT $-$ o & 0.71 ${\color{darkred}\pm}$ 0.09 & 0.76 ${\color{darkred}\pm}$ 0.08 &  2.08 ${\color{darkred}\pm}$ 0.30 & 1.75 ${\color{darkred}\pm}$ 0.16 & -1.82 ${\color{darkred}\pm}$ 0.46 & -1.70 ${\color{darkred}\pm}$ 0.21 \\\\

HS  & 0.09 ${\color{darkred}\pm}$ 0.03 & 0.18 ${\color{darkred}\pm}$ 0.02 &  1.31 ${\color{darkred}\pm}$ 0.37 & 1.30 ${\color{darkred}\pm}$ 0.11 & -2.25 ${\color{darkred}\pm}$ 0.64 & -1.80 ${\color{darkred}\pm}$ 0.25 \\\\

HS $-$ o & 0.09 ${\color{darkred}\pm}$ 0.03 & 1.02 ${\color{darkred}\pm}$ 0.08 &  1.31 ${\color{darkred}\pm}$ 0.37 & 1.54 ${\color{darkred}\pm}$ 0.13 & -2.28 ${\color{darkred}\pm}$ 0.66 & -1.80 ${\color{darkred}\pm}$ 0.25 \\\\

%TT & 0.80 ${\color{darkred}\pm}$ 0.12 & 0.05 ${\color{darkred}\pm}$ 0.01 &  4.20 ${\color{darkred}\pm}$ 0.52 & 1.82 ${\color{darkred}\pm}$ 0.19 & -1.45 ${\color{darkred}\pm}$ 0.31 & -1.77 ${\color{darkred}\pm}$ 0.30 \\\\

%TT $-$ o & 3.31 ${\color{darkred}\pm}$ 0.29 & 1.46 ${\color{darkred}\pm}$ 0.11 &  4.33 ${\color{darkred}\pm}$ 0.53 & 2.04 ${\color{darkred}\pm}$ 0.22 & -1.44 ${\color{darkred}\pm}$ 0.31 & -1.77 ${\color{darkred}\pm}$ 0.30 \\\\

%CJT & 0.08 ${\color{darkred}\pm}$ 0.03 & 0.18 ${\color{darkred}\pm}$ 0.02 &  1.31 ${\color{darkred}\pm}$ 0.37 & 1.31 ${\color{darkred}\pm}$ 0.11 & -2.04 ${\color{darkred}\pm}$ 0.51 & -1.62 ${\color{darkred}\pm}$ 0.19 \\\\

%CJT $-$ o & 0.08 ${\color{darkred}\pm}$ 0.03 & 1.01 ${\color{darkred}\pm}$ 0.08 &  1.31 ${\color{darkred}\pm}$ 0.37 & 1.55 ${\color{darkred}\pm}$ 0.13 & -2.04 ${\color{darkred}\pm}$ 0.51 & -1.62 ${\color{darkred}\pm}$ 0.19 \\\\

\end{tabular}
}
\begin{tablenotes}\footnotesize \small \vspace{+1mm}
\item {\color{blue}$\ast$} \hspace{0.5cm}  $-$ o are transitions without taking account the outlier XTE 1550-564 in 1999 outburst.
\item {\color{blue}a} \hspace{0.5cm} weight = $1 / (\sigma)^2$
\item {\color{blue}b} \hspace{0.5cm} weight = $(\rm value / \sigma)^2$
\item {\color{blue}c} \hspace{0.5cm} in log$_{10}$
\end{tablenotes}

\end{table*}

\vspace{1cm}
\begin{table*}
\centering
\caption[]{State transition ELF of sources with good mass and distance measurement (4U 1543-47, GRO J1655-40, GX 339-4, H1743-322, XTE 1550-564).}\label{tab:wmean3}
\scalebox{0.95}{
\begin{tabular}{l|cc|cc|cc}
\hline \hline
 & \multicolumn{2}{c}{Weighted mean 1} & \multicolumn{2}{c}{Weighted mean 2} & \multicolumn{2}{c}{Histogram} \Tstrut\Bstrut \\
Transition type  &  DBB & PLF & DBB & PLF & DBB & PLF \\
 & (\%) & (\%) & (\%) & (\%) & (\%) & (\%) \\
\hline \Tstrut  \\
SIMS & 2.79 ${\color{darkred}\pm}$ 0.50 & 1.88 ${\color{darkred}\pm}$ 0.35 &  2.94 ${\color{darkred}\pm}$ 0.53 & 2.79 ${\color{darkred}\pm}$ 0.53 &  -- & -- \\\\

SIMS $-$ o & 2.79 ${\color{darkred}\pm}$ 0.50 & 1.88 ${\color{darkred}\pm}$ 0.35 &  2.94 ${\color{darkred}\pm}$ 0.53 & 2.79 ${\color{darkred}\pm}$ 0.52 &  -- & -- \\\\

HIMS  & 0.74 ${\color{darkred}\pm}$ 0.12 & 0.05 ${\color{darkred}\pm}$ 0.01 &  4.09 ${\color{darkred}\pm}$ 0.54 & 1.90 ${\color{darkred}\pm}$ 0.22 &  -1.47 ${\color{darkred}\pm}$ 0.29 & -1.75 ${\color{darkred}\pm}$ 0.29
 \\\\
 
HIMS $-$ o & 1.95 ${\color{darkred}\pm}$ 0.23 & 1.13 ${\color{darkred}\pm}$ 0.13 &  4.23 ${\color{darkred}\pm}$ 0.56 & 2.19 ${\color{darkred}\pm}$ 0.26 & -1.48 ${\color{darkred}\pm}$ 0.29 & -1.75 ${\color{darkred}\pm}$ 0.29 \\\\

IT  & 0.32 ${\color{darkred}\pm}$ 0.05 & 0.19 ${\color{darkred}\pm}$ 0.03 &  1.67 ${\color{darkred}\pm}$ 0.25 & 1.72 ${\color{darkred}\pm}$ 0.16 & -1.88 ${\color{darkred}\pm}$ 0.34 & -1.71 ${\color{darkred}\pm}$ 0.16 \\\\

IT $-$ o & 0.68 ${\color{darkred}\pm}$ 0.09 & 1.57 ${\color{darkred}\pm}$ 0.14 &  1.70 ${\color{darkred}\pm}$ 0.26 & 1.82 ${\color{darkred}\pm}$ 0.17 & -1.86 ${\color{darkred}\pm}$ 0.33 & -1.71 ${\color{darkred}\pm}$ 0.16 \\\\

HS  & 0.26 ${\color{darkred}\pm}$ 0.07 & 0.18 ${\color{darkred}\pm}$ 0.02 &  0.79 ${\color{darkred}\pm}$ 0.18 & 1.19 ${\color{darkred}\pm}$ 0.10 & -2.42 ${\color{darkred}\pm}$ 0.40 & -1.80 ${\color{darkred}\pm}$ 0.20 \\\\

HS $-$ o & 0.26 ${\color{darkred}\pm}$ 0.07 & 1.14 ${\color{darkred}\pm}$ 0.10 &  0.79 ${\color{darkred}\pm}$ 0.18 & 1.43 ${\color{darkred}\pm}$ 0.13 & -2.45 ${\color{darkred}\pm}$ 0.42 & -1.80 ${\color{darkred}\pm}$ 0.20 \\\\

%TT & 0.76 ${\color{darkred}\pm}$ 0.12 & 0.05 ${\color{darkred}\pm}$ 0.01 &  3.76 ${\color{darkred}\pm}$ 0.49 & 1.60 ${\color{darkred}\pm}$ 0.19 & -1.38 ${\color{darkred}\pm}$ 0.18 & -1.65 ${\color{darkred}\pm}$ 0.14 \\\\

%TT $-$ o & 2.03 ${\color{darkred}\pm}$ 0.23 & 1.11 ${\color{darkred}\pm}$ 0.13 &  3.89 ${\color{darkred}\pm}$ 0.51 & 1.86 ${\color{darkred}\pm}$ 0.22 & -1.38 ${\color{darkred}\pm}$ 0.18 & -1.65 ${\color{darkred}\pm}$ 0.14 \\\\

%CJT & 0.26 ${\color{darkred}\pm}$ 0.07 & 0.17 ${\color{darkred}\pm}$ 0.02 &  0.79 ${\color{darkred}\pm}$ 0.18 & 1.22 ${\color{darkred}\pm}$ 0.11 & -2.25 ${\color{darkred}\pm}$ 0.26 & -1.65 ${\color{darkred}\pm}$ 0.19 \\\\

%CJT $-$ o & 0.26 ${\color{darkred}\pm}$ 0.07 & 1.15 ${\color{darkred}\pm}$ 0.10 &  0.79 ${\color{darkred}\pm}$ 0.17 & 1.48 ${\color{darkred}\pm}$ 0.13 & -2.25 ${\color{darkred}\pm}$ 0.26 & -1.65 ${\color{darkred}\pm}$ 0.19 \\\\

\end{tabular}
}
%\begin{tablenotes}\footnotesize \small \vspace{+1mm}
%\item {\color{blue}$\ast$} \hspace{0.5cm}  $-$ o are transitions without taking account the outlier XTE 1550-564 in 1999 outburst.
%\item {\color{blue}a} \hspace{0.5cm} weight = $1 / (\sigma)^2$
%\item {\color{blue}b} \hspace{0.5cm} weight = $(\rm value / \sigma)^2$
%\item {\color{blue}c} \hspace{0.5cm} in log$_{10}$
%\end{tablenotes}

\end{table*}

\vspace{1cm}
\begin{table*}
\centering
\caption[]{State transition ELF of sources with poor mass and distance measurement (XTE 1650-500, XTE 1720-318, XTE 1748-288, XTE 2012+381, XTE 1817-330, XTE 1908+094, XTE 1752-223).}\label{tab:wmean4}
\scalebox{0.95}{
\begin{tabular}{l|cc|cc}
\hline \hline & \multicolumn{2}{c}{Weighted mean 1} 
 & \multicolumn{2}{c}{Weighted mean 2}  \Tstrut\Bstrut \\ Transition type  &  DBB & PLF & DBB & PLF \\
 & (\%) & (\%) & (\%) & (\%)  \\
\hline \Tstrut  \\

%SIMS & -- ${\color{darkred}\pm}$ -- & -- ${\color{darkred}\pm}$ -- &  -- ${\color{darkred}\pm}$ -- & -- ${\color{darkred}\pm}$ --\\\\

HIMS  & 2.31 ${\color{darkred}\pm}$ 0.76 & 0.60 ${\color{darkred}\pm}$ 0.12 &  5.36 ${\color{darkred}\pm}$ 1.45 & 2.87 ${\color{darkred}\pm}$ 0.42 \\\\
 
IT  & 2.76 ${\color{darkred}\pm}$ 0.87 & 0.47 ${\color{darkred}\pm}$ 0.10 &  3.34 ${\color{darkred}\pm}$ 1.26 & 2.07 ${\color{darkred}\pm}$ 0.34 \\\\

HS  & 0.05 ${\color{darkred}\pm}$ 0.04 & 0.90 ${\color{darkred}\pm}$ 0.16 &  2.30 ${\color{darkred}\pm}$ 1.31 & 2.08 ${\color{darkred}\pm}$ 0.27 \\\\

%TT & 2.61 ${\color{darkred}\pm}$ 0.87 & 0.81 ${\color{darkred}\pm}$ 0.14 &  7.18 ${\color{darkred}\pm}$ 2.27 & 2.44 ${\color{darkred}\pm}$ 0.37 \\\\

%CJT & 0.03 ${\color{darkred}\pm}$ 0.04 & 0.87 ${\color{darkred}\pm}$ 0.15 &  3.20 ${\color{darkred}\pm}$ 1.61 & 2.24 ${\color{darkred}\pm}$ 0.34 \\\\
\end{tabular}
}
%\begin{tablenotes}\footnotesize \small \vspace{+1mm}
%\item {\color{blue}a} \hspace{0.5cm} weight = $1 / (\sigma)^2$
%\item {\color{blue}b} \hspace{0.5cm} weight = $(\rm value / \sigma)^2$
%\end{tablenotes}

\end{table*}

%% /*******************************************************************
%% ** Results - Arithmetic and weighted mean                                                        **
%% *******************************************************************/

\subsection{Weighted mean of state transition luminosities}\label{sec:avmean}

We first investigated disk, power, and total state transition luminosity distributions for all outbursts and transition types by just plotting luminosities of all outbursts in three panels: DBB, PL and total ELF. Two examples are given in Figures~\ref{fig:trans_hims} and \ref{fig:trans_hs}. This exercise provided us information on the transitions that would result in the least scattering around the mean, and individual outbursts that deviate most from the mean. From Figures~\ref{fig:trans_hims} and \ref{fig:trans_hs} it is immediately obvious that XTE~J1550$-$564 in 1999 outburst decay exhibits state transitions at much lower luminosities compared to other outbursts. We discuss this particular outburst in \S~\ref{subsec:outlier}. We then calculated the weighted means of the transition luminosities and their corresponding errors for all outbursts and transitions to quantify the properties of state transition luminosity distributions. To calculate the error in the mean, we first grouped transition fluxes source-by-source, then calculated the relative error in flux assuming they are uncorrelated for each source, and finally added the error coming from the mass and distance to the mean flux error in quadratics to obtain the overall error per source in luminosity. 
With this method, we avoided treating correlated mass and distance errors for the overall distribution.

The luminosities were first weighted by the inverse squares of their standard deviation (\citealt{maccarone2003x}, \autoref{tab:wmean}, first column). However, this weighting method is strongly affected by the anomalously low luminosity outburst of XTE J1550$-$564 since all error components in flux, mass, and distance are small, resulting in a very large weight for this outburst. We calculated weighted means by both including and excluding this outburst only and reported all results in \autoref{tab:wmean} (rows indicated with - o). During the transition to the hard state, the power-law luminosity has a weighted mean of 1.02 $\pm$ 0.08 ELF excluding and 0.18 $\pm$ 0.02 ELF including this outburst. Similarly, during the transition to HIMS, including this outburst to the mean decreases the average from  0.79 $\pm$ 0.09 ELF to 0.05 $\pm$ 0.01 ELF. Since directly weighting with the square of the absolute error results in a single observation to dominate the mean overall observations, we also tried weighting with squares of the relative errors rather than the absolute errors (\autoref{tab:wmean}, second column). With this weighting method, the power-law luminosity has a weighted mean of 1.30 $\pm$ 0.11 ELF excluding and 1.54 $\pm$ 0.13 ELF including 1999 outburst of XTE J1550$-$564 during the transition to hard state. 

There are only 4 outbursts with transitions out of SIMS, therefore we remove the SIMS transition out of the discussion, but provide values here for possible future studies. Except for the SIMS transition, the errors in DBB are large, 10-80\%, which is not surprising given the poor low energy response of the PCA instrument. In general, an unweighted average of all power law luminosities results in a state transition luminosity of around 2\% with around 20\% uncertainty.

\subsection{Histograms}\label{sec:histogram}

\begin{figure}
\includegraphics[width=85mm]{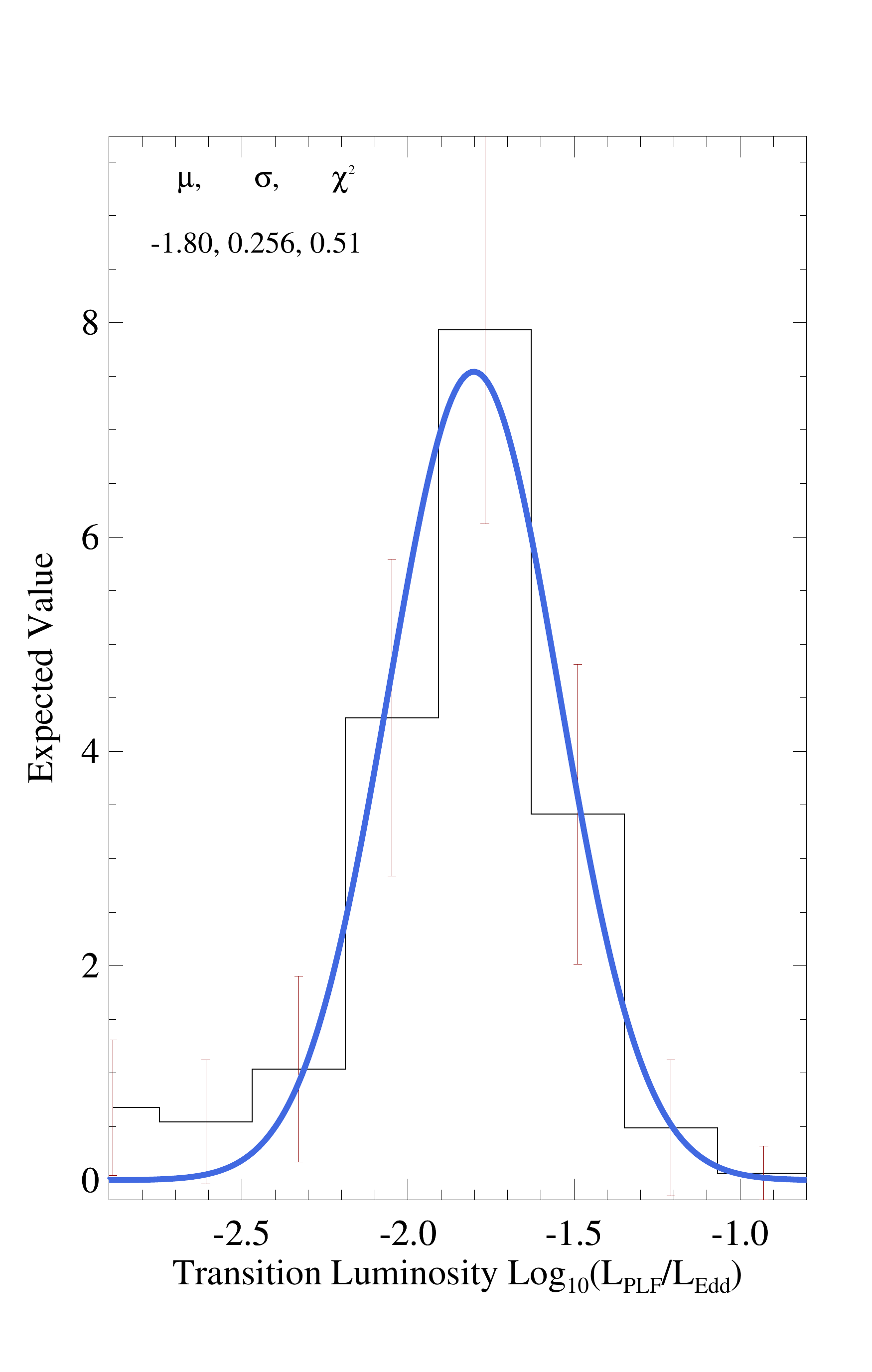}
\vspace{-1.30 cm}
\caption{\label{fig:hs_pl_hist}
A fit to the distribution of state transition luminosity of PL during the transition to the hard state after accounting for the $x-$axis errors using the Monte Carlo simulations described in the text.  The mean value of the log ELF, its standard deviation, and the reduced \textchi$^2$ are given in the upper left.  The linear value of the best fitting model is 1.58 $\pm$ 0.93 ELF.}
\end{figure}

\begin{figure}
\includegraphics[width=85mm]{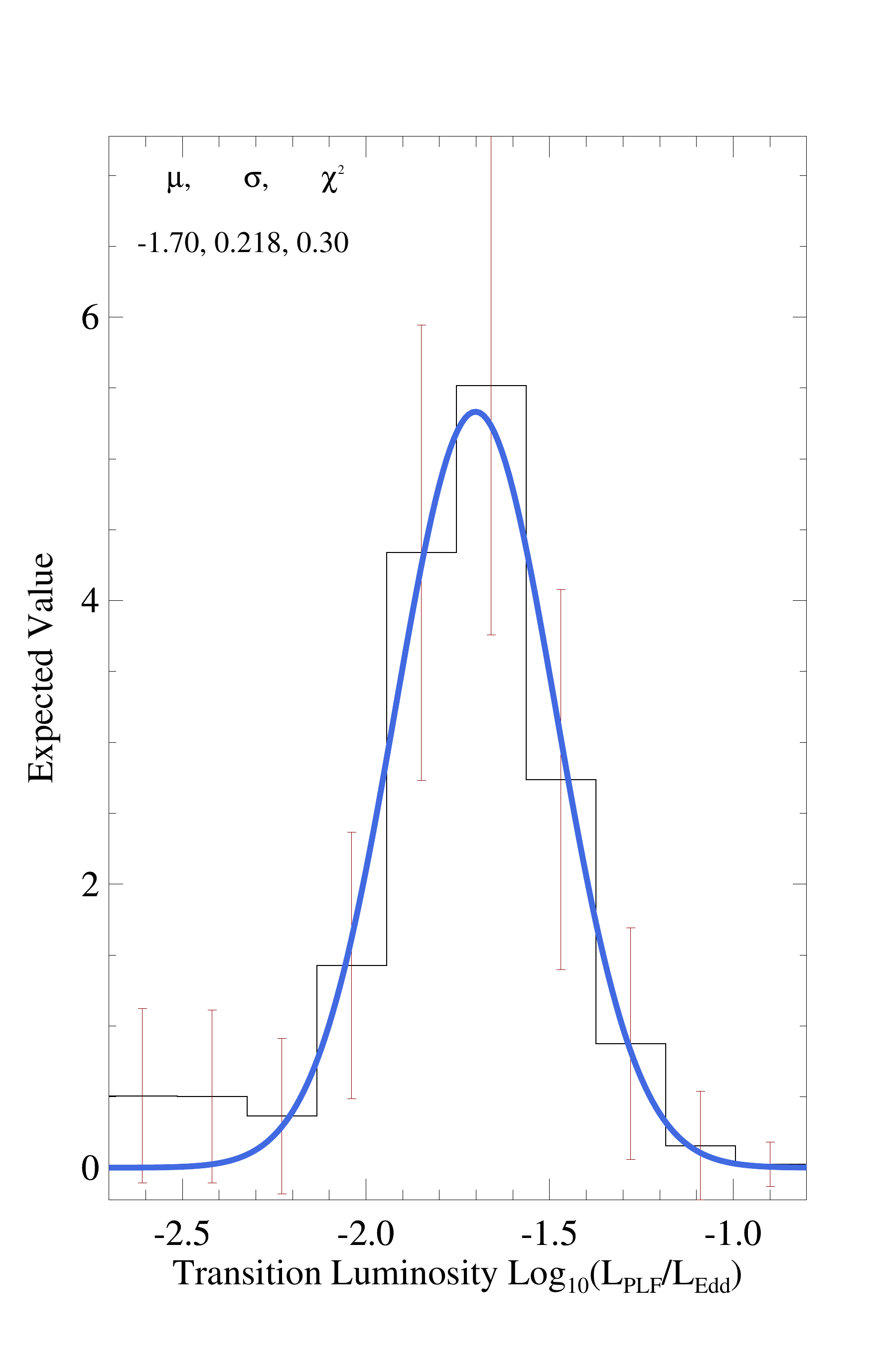}
\vspace{-1.30 cm}
\caption{\label{fig:it_pl_hist}
A fit to the distribution of state transition luminosity of PL during the index transition after accounting for the $x-$axis errors using the Monte Carlo simulations described in the text.  The mean value of the log ELF, its standard deviation, and the reduced \textchi$^2$ are given in the upper left.  The linear value of the best fitting model is 1.99 $\pm$ 1.00 ELF.}
\end{figure}

\begin{figure}
\includegraphics[width=85mm]{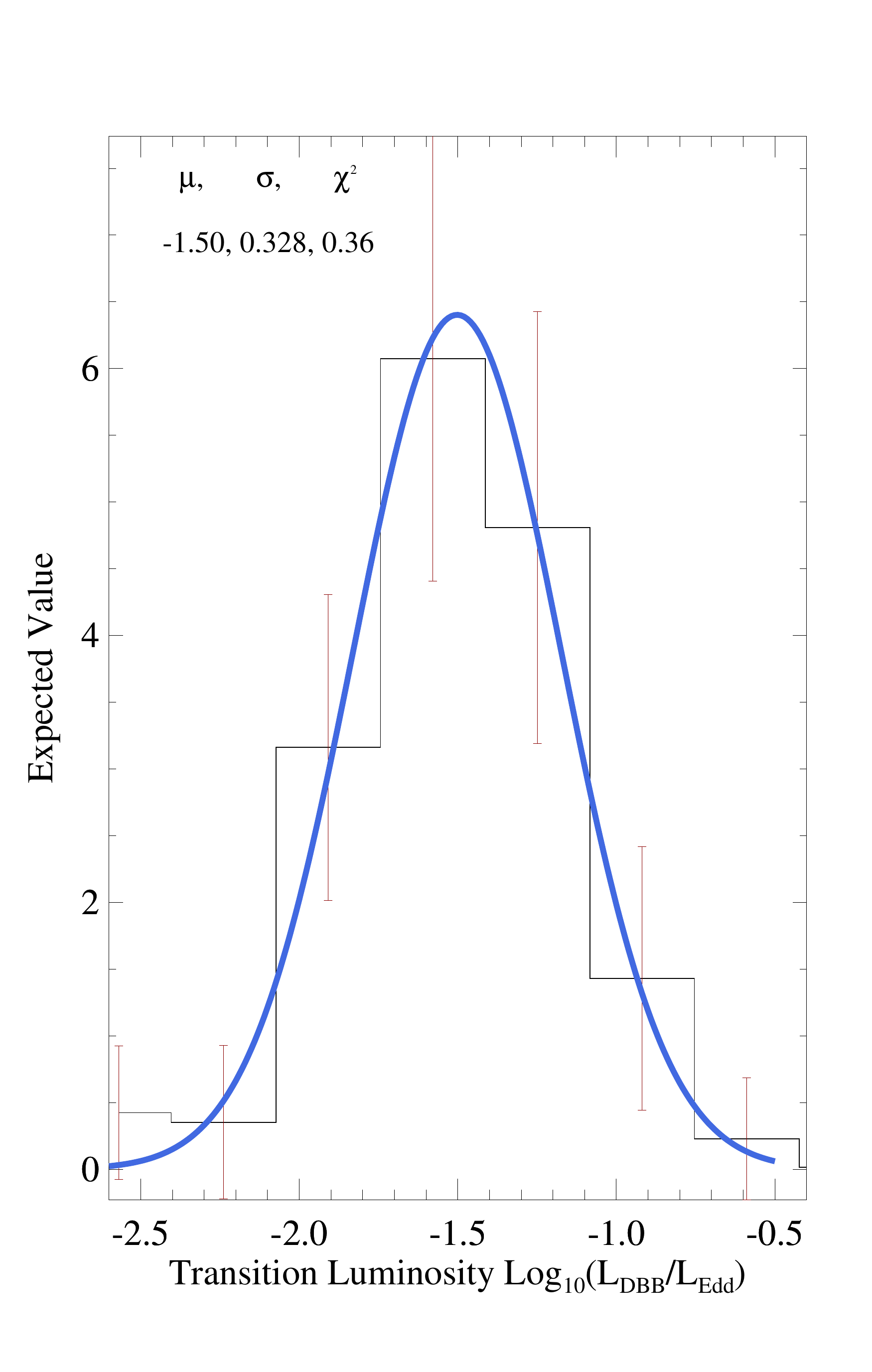}
\vspace{-1.30 cm}
\caption{\label{fig:hims_hist}
A fit to the distribution of state transition luminosity of DBB during the transition to HIMS after accounting for the $x-$axis errors using the Monte Carlo simulations described in the text.  The mean value of the log ELF, its standard deviation, and the reduced \textchi$^2$ are given in the upper left.  The linear value of the best fitting model is 3.16 $\pm$ 2.38 ELF.}
\end{figure}

Since the data from a  single outburst may cause a large deviation in the weighted means, we have also tried fitting histograms of the state transition luminosities. We have performed Monte Carlo simulations in order to take into account the errors on the x-axis by randomly selecting a state transition luminosity value from a Gaussian distribution with 1$\sigma$ values retrieved from the propagated error in the luminosity calculation similar to the methodology described in \cite{dunn2010global} and \cite{Tetarenko2016ApJS}. We then fitted a Gaussian distribution to our expected distribution and obtained the mean, sigma, and reduced chi-square. We have repeated this procedure for each state transition of interest and presented the results in \autoref{tab:wmean}. With this method, we obtained a log fraction transition luminosity and an error of -1.80 $\pm$ 0.25 log ELF (1.58 $\pm$ 0.93 ELF) for power-law during the transition to the hard state (\autoref{fig:hs_pl_hist}). As it can be seen in \autoref{fig:it_pl_hist},  we have obtained even a tighter clustering during the index transition with a log luminosity and an error of -1.70 $\pm$ 0.21 (1.99 $\pm$ 1.00 ELF) for the power-law. Similarly, in \autoref{fig:trans_hims}, we have found a clustering for the disk emission during the transition to HIMS with a log luminosity and an error of -1.50 $\pm$ 0.32 (3.16 $\pm$ 2.38 ELF) (\autoref{fig:hims_hist}).

%% /*******************************************************************
%% ** Discussion                                                        **
%% *******************************************************************/

\section{Discussion}\label{sec:discussion}

We have calculated state transition luminosities of GBHTs in outburst decay and noticed  different clustering behavior for power-law and thermal disk emission at different transitions. Here, we first compare our results with those that exist in the literature.
 
 The first claim of approximately constant state transition luminosity during the outburst decay was made in \cite{maccarone2003x}. The majority of the state transition luminosities were calculated using fluxes in literature and with a crude bolometric correction without taking into account the type of transition or the exact spectral shape; yet the average state transition luminosity found in this work (1.9 $\pm$ 0.2 \% $L_{\rm Edd}$) is quite consistent with our results for the power-law luminosity transition to the hard state. \cite{maccarone2003x} also argues that good temporal coverage is more important than good broadband spectral coverage as errors in bolometric corrections are small if the mass and distance are well determined. For the transition to the hard state we would agree with this assessment; however, for the index transition or the earlier transitions, bolometric corrections become much more critical, especially for obtaining the DBB luminosities.
 
\cite{dunn2010global} used a Monte Carlo simulation for sources with unknown mass and distances by randomly picking masses and distances from a comparably large range of mass M = 10 $\pm$ 10 M$_{\odot}$ and distance d = 5 $\pm$ 5 kpc distributions many times to create a histogram of state transition luminosities. Although Their methodology has led to broader distribution (log ELF = -1.57 $\pm$ 0.59), their soft to hard transition luminosities is consistent with our results for the PL luminosity.

\cite{Tetarenko2016ApJS} have calculated both the fluxes and the bolometric luminosities of the sources by assuming a spectrum consisting of a soft DBB and a hard Comptonized component. They have adopted a Markov chain Monte Carlo (MCMC) method in order to find the ratio of hard to soft band flux density alongside with normalization parameters with 1$\sigma$ confidence intervals. For the sources without distance measurements, they have adopted a uniform distribution between 3 kpc and 8 kpc and sampled from that distribution. Their soft to hard transition luminosities (log ELF = -1.50 $\pm$ 0.37) is also consistent with our findings for the PL luminosity during the transition to the hard state.

In this study, for each observation, we have applied spectral and temporal analyses in order to accurately determine state transitions. Furthermore, we have separated disk and power-law luminosity fractions which not only allowed us to compare the luminosity distribution between these two components in different states but also their correlation with other parameters such as the inclination. Finally, for the sources without a distance measurement, we implemented a distance estimation based on the high likelihood that the sources will be found in Galactic arms and the bulge, and possibly be at a distance greater than 4 kpc. This approach has led us to observe a narrow distribution in power-law luminosity during the transition to hard state and find a second and tighter clustering during index transition as well as a third clustering in DBB luminosity during the transition to HIMS.

\subsection{Outliers}\label{subsec:outlier}

\subsubsection{4U 1543-47}

The transition luminosity of the DBB component of 4U 1543-47 is found to be high compared to those of the other sources, in almost all states. This is not the case for its power-law transition luminosities.  This difference is most likely  due to its lower inclination angle \citep{2006MNRASGierlinski}. This would make sense if an isotropic emission is assumed for the Comptonized region (see \ref{subsec:inclination}), since then the disk emission would be enhanced due to being face-on, but the power law emission would not be. 

\subsubsection{XTE J1550-564}

The transitions in the decay phase of 1999 outburst of XTE J1550-564 occur at very low luminosities, both the power-law and the DBB luminosities are an order of magnitude lower than the general trend.

\cite{Sobczak2000ApJS} have covered complete spectral analyses of the source in 1998-1999 outburst and found that between MJD 51240 and MJD 51253, the DBB has dropped and power-law flux has increased as the source entered SIMS. If this decaying trend had continued, the source could have gone into the hard state at luminosities similar to observed during the 2000 outburst. However, the trend reversed and the DBB increased and dominated the spectrum for another 20 days.  While \cite{Sobczak2000ApJS} interpreted this behavior as a hard X-ray flare which possibly occurred when the source was in the intermediate state, another interpretation could be that the outburst decay has been interrupted by a new mass flow reigniting the soft emission. In a sense, a secondary outburst happened pushing the source into a new soft state at a lower luminosity than usual, thereby its decay to the hard state occurred at a very low luminosity level.

\subsubsection{XTE J1908+094}

XTE J1908+094 is another transient source that its state transition luminosity is below average during outburst decay. Its PL luminosity is up to 50\% less than the weighted mean during different state transitions. The disk also showed a steep decay and went below the detection limit just before the transition to HIMS which led to an average total luminosity of 0.33 ELF during the transition to HIMS as well as 0.4 ELF during the transition to the hard state. The apparently low state transition luminosity of XTE J1908+094 is a different case.  For this object, the black hole mass has not been estimated dynamically, and the distance use is the default $8 \pm 2$ kpc used for sources with no other distance information.  It is thus of real importance to understand the mass and distance for this object to determine whether it is an outlier, has an especially low mass black hole, or is simply located further away.  Notably, the Norma spiral arm of the Galaxy is at a distance of 13.5 kpc, and if the source is associated with the Norma arm, it could have a \say{standard} transition luminosity while having a standard black hole mass.

\subsubsection{4U 1630-47}

As it mentioned in \S\ref{sec:datareduction},  4U 1630-47 were not included in the analyses due to its peculiar behaviors. In the majority of outbursts, the source did not follow the typical hysteresis pattern observed in the HID. \cite{Tomsick2014ApJT} have reported a delayed transition to the hard state for 2010 outburst of 4U 1630$-$47 where spectrum remained soft until $\sim$ 50 days after the main outburst ended. The presence of these outlier cases would suggest the idea of unknown accretion flow characteristics.

\subsection{Inclination angle}\label{subsec:inclination}

\begin{figure}
\includegraphics[width=85mm]{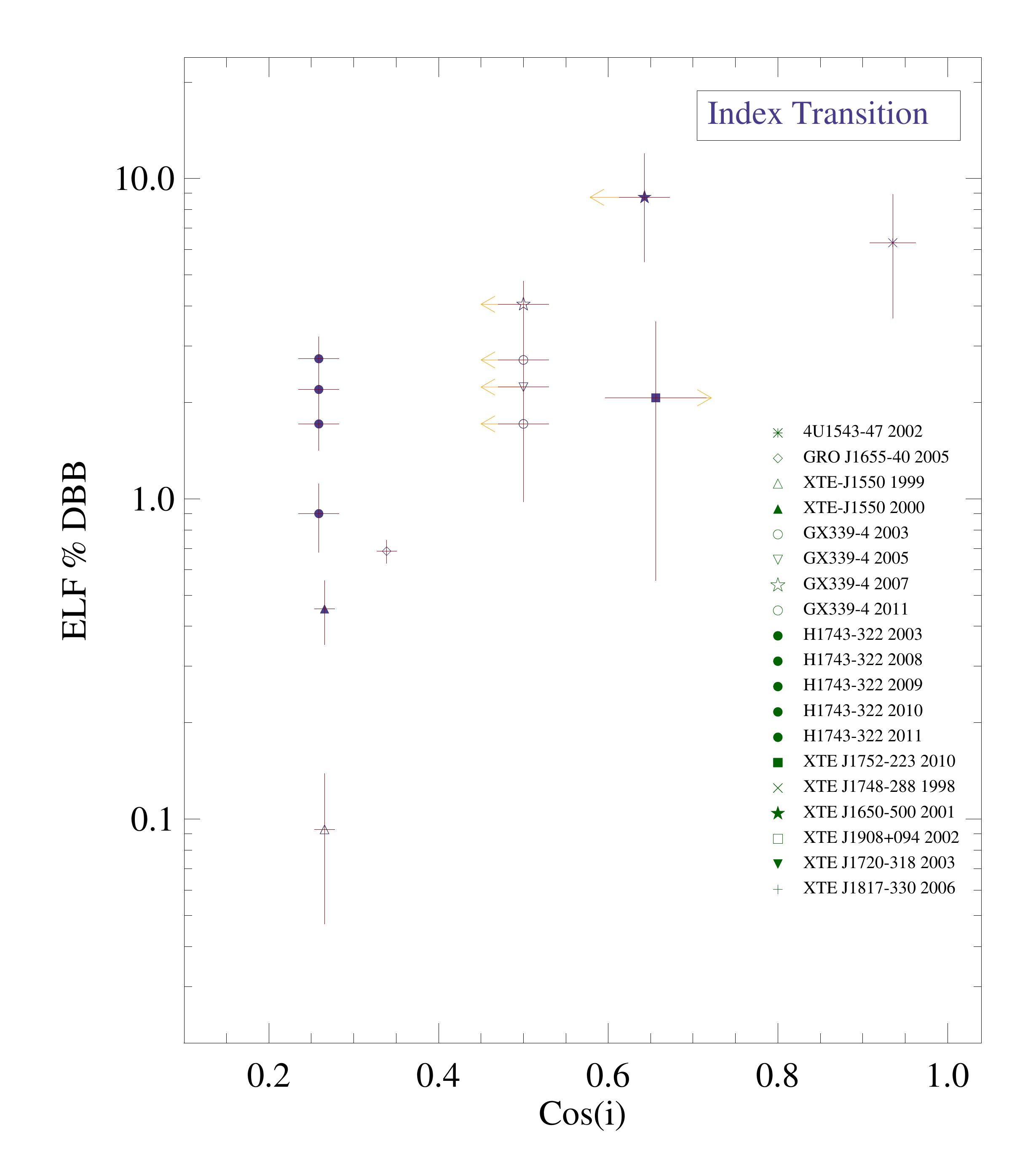}
\vspace{-0.50 cm}
\caption{\label{fig:inc_dbb}
ELF of DBB luminosity as a function of cosine of the inclination angle during index transition. The corresponding references are given in \autoref{tab:bhtransients}.}
\end{figure}

\begin{figure}
\includegraphics[width=85mm]{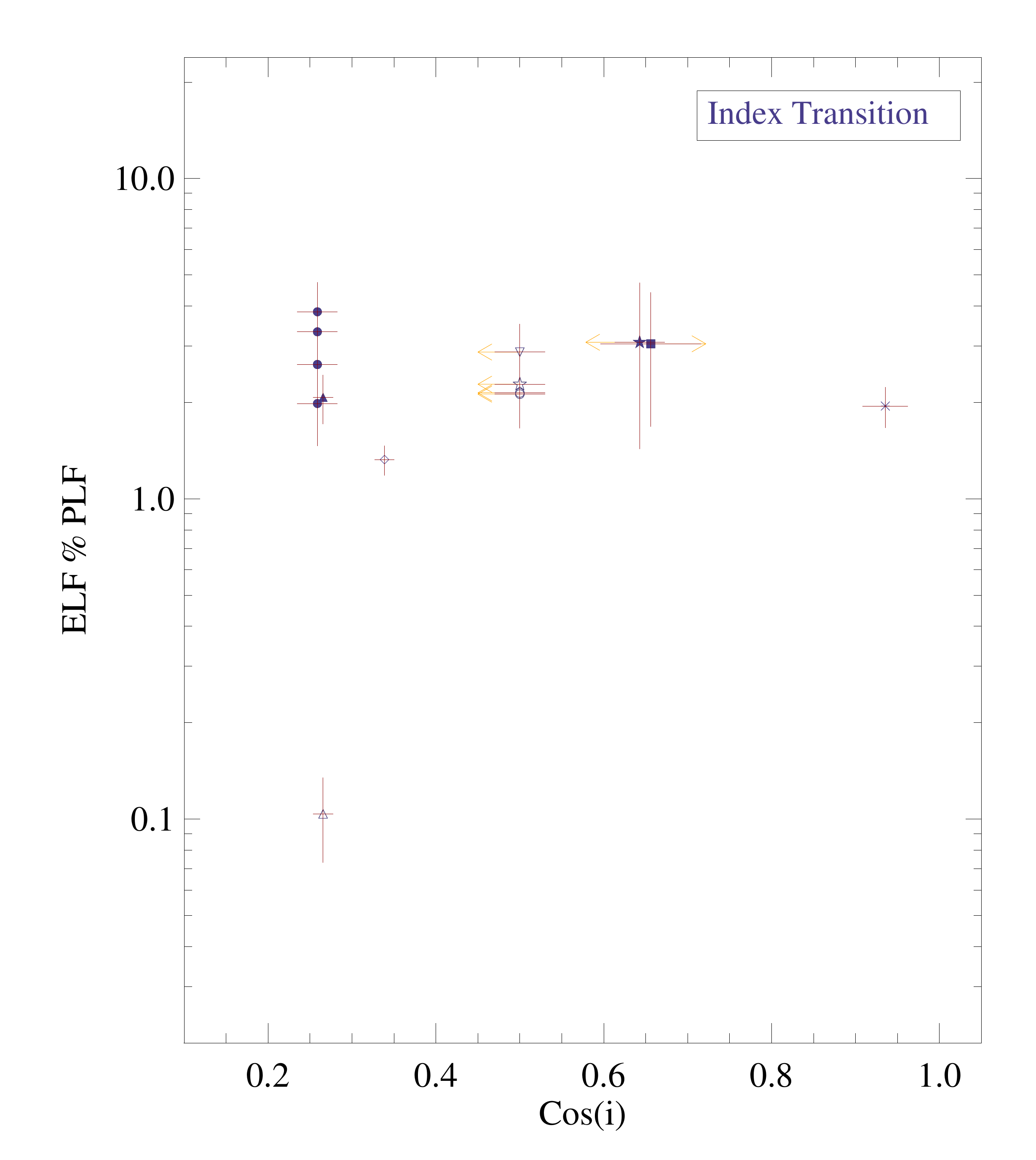}
\vspace{-0.50 cm}
\caption{\label{fig:inc_pl}
ELF of PL luminosity as a function of cosine of the inclination angle during index transition. The corresponding references are given in \autoref{tab:bhtransients}.}
\end{figure}

We have plotted Eddington luminosity fractions for both power-law and DBB emission in all state transitions as a function of inclination angles (see \autoref{tab:bhtransients} for the inclination values). For power-law emission, we have observed a flat distribution in almost all state transitions (\autoref{fig:inc_pl}) which indicates that there are no correlations between these two parameters, the emission is isotropic. On the other hand, for the majority of state transitions, we have obtained a positive correlation between the DBB luminosity and cosine of the inclination angle (\autoref{fig:inc_dbb}) with a median Pearson's rank correlation coefficient of 0.73. We concluded that for thermal disk emission, an inclination angle correction is necessary in order to compare state transition luminosities of different sources. This could at least partially explain the wider transition luminosity distributions seen in \cite{dunn2010global} and \cite{Tetarenko2016ApJS}.

Furthermore, this result might give some hint regarding the possible geometry and size of the corona as well as reformation process. The standard accretion disk coronae model \citep{Bisnovatyi-Kogan1977A&A,Galeev1979ApJ} as well as advection \citep{Narayan1994ApJ} and convective \citep{Stone1996ApJ} dominated accretion flow (ADAF \& CDAF) models. The latter models often invoke hole in the middle type of corona geometry. The accretion disk corona, on the other hand, may have a ``sandwich'' or ``slab'' geometry.  
\cite{Dove1997ApJ} already showed that in the case of the slab or full sandwich geometry, for a given total optical depth ($\tau$) of the planar corona, there is no self-consistent coronal temperature that can produce a hard spectrum with a high exponential cut off observed in Cyg X-1. However, this argument is not sufficient to rule out other geometries such as patchy corona where the majority of photons reprocessed in the disk are able to escape without returning to the source, therefore, reducing the cooling rate.

The high bulk velocity may have been responsible for the strong anisotropy observed in these systems. Such velocity may have arisen either from a bulk motion comptonization (BMC) in a very rapidly rotating case or from an outflowing corona. 
In the hole in the middle type hot flow models, if synchrotron radiation is the source of the seed photons, then the emission should be almost isotropic since there is not really a strongly preferred angle in this model.

If the majority of seed photons originates from the inner part of the accretion disk, then the photons have a preferred starting direction due to the geometry of the disk. However,  since Compton up-scattering is a nearly isotropic process, the anisotropy effects should be negligible. In the electron rest frame, there is a cos$^2$ dependency in the cross-section, however, since electrons most likely have random directions of motion in the hot flow, after the frame transformations the emission should not have any preferred direction. Finally, The Comptonization process predicts that the backscattered electrons get the highest boost in energy whereas the forwardly scattered electrons get almost no boost and the ones scattered by 90 degrees get intermediate boosts. Therefore, this effect would almost cancel out in total considering the backscattering is suppressed a bit due to the Klein-Nishina limit. All of these effects will be minor as the flow becomes optically thick, but the probability of escaping out the plane of the disk would increase. We predict that for the optical depth less than one, the effect of inclination angle are  insignificant unless there is a relativistic bulk outflow. Given that the sample of inclination angles is small, and that the inclination angles come from a heterogeneous set of methods, we can only cautiously interpret this correlation, but it is in agreement with theoretical expectations for the behaviour of the thermal component and highlights the need for a larger sample of good inclination angle measurements.

\section{Summary and Conclusion}

In this study, by using the large sample of GBHTs from the RXTE archival data, we have investigated the state transition luminosity distribution of sources in the outburst decay by separating power-law and disk components. By performing both spectral and temporal analyses, we were able to distinguish several state transitions reported in \cite{Belloni10_jp} and \cite{kalemci2013complete}. Our results can be summarized as follows:

\begin{itemize}
 \setlength\itemsep{0.8em}
    \item For the PL flux, we have obtained a tight clustering with a luminosity and the error of -1.70 $\pm$ 0.21 log ELF during the index transition and -1.80 $\pm$ 0.25 log ELF during the transition to the hard state. 
    \item For the DBB flux, we have found another clustering with a luminosity and the error of -1.50 $\pm$ 0.32 log ELF during the transition to HIMS.
    \item We have interpreted the low transition luminosity of 1999 outburst of XTE J1550-564 as a disruption of a new mass flow and transition to a low luminous soft state (a secondary outburst).
    \item We have found that only the DBB luminosity correlates with the cosine of the inclination angle whereas PL luminosity showed a flat distribution in almost all state transitions. This would not only highlight the importance of the inclination angle correction in luminosity calculations, but it might also be suggesting a spherical, hole in the middle type geometry for the hot-flow. 
\end{itemize}

%% /*******************************************************************
%% ** Acknowledgments                                               **
%% *******************************************************************/

\section*{Acknowledgements}
 We thank the anonymous referee for constructive comments that helped to improve the manuscript. A.VM and E.K acknowledge support from T\"UB\.ITAK 1001 project 115F488. A.VM acknowledge support from Istanbul Technical University Scientific Research Projects Unit (ITU-BAP) project 40196. A. VM thank Eda Vurgun for support with figures. This research has made use of data, software, and/or web tools obtained from the High Energy Astrophysics Science Archive Research Center, provided by the NASA/Goddard Space Flight Center.
 
%% /*******************************************************************
%% ** Bibliography                                                   **
%% *******************************************************************/

\bibliographystyle{mn2e_fixed}
%\bibliography{refs}

\clearpage
%%%%%%%%%%%%%%%%%%%%%%%%%%%%%%%%%%%%%%%%%%%%%%%%%%

%%%%%%%%%%%%%%%%% APPENDICES %%%%%%%%%%%%%%%%%%%%%

%\clearpage
\appendix

\section{Hardness intensity diagram of GBHT\lowercase{s}}

\begin{figure}
\includegraphics[width=85mm]{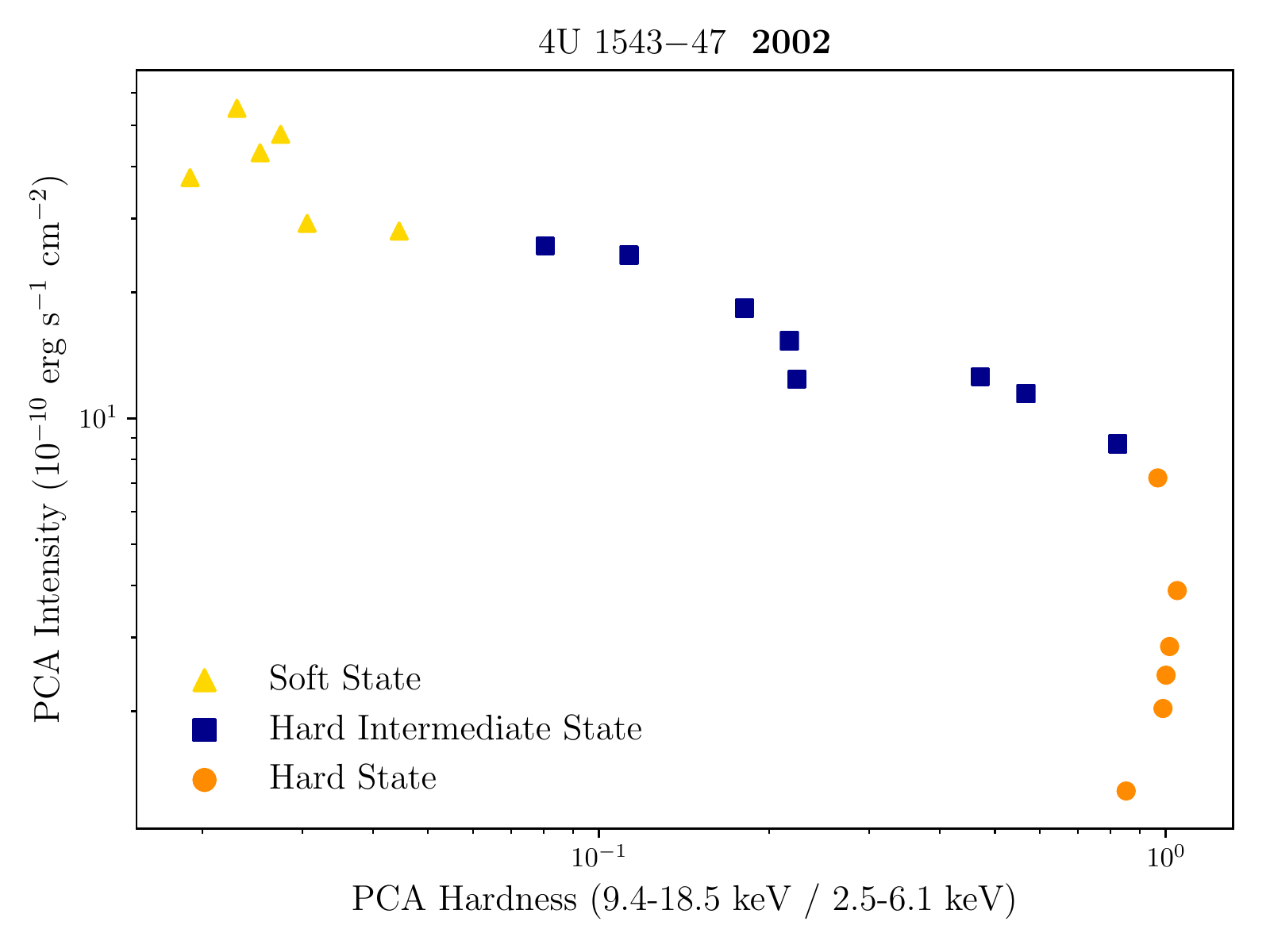}
\vspace{-0.50 cm}
\caption{\label{fig:4U15432002_HID}
Hardness intensity diagram of 4U 1543$-$47 in 2002 outburst decay.}
\end{figure}

\begin{figure}
\includegraphics[width=85mm]{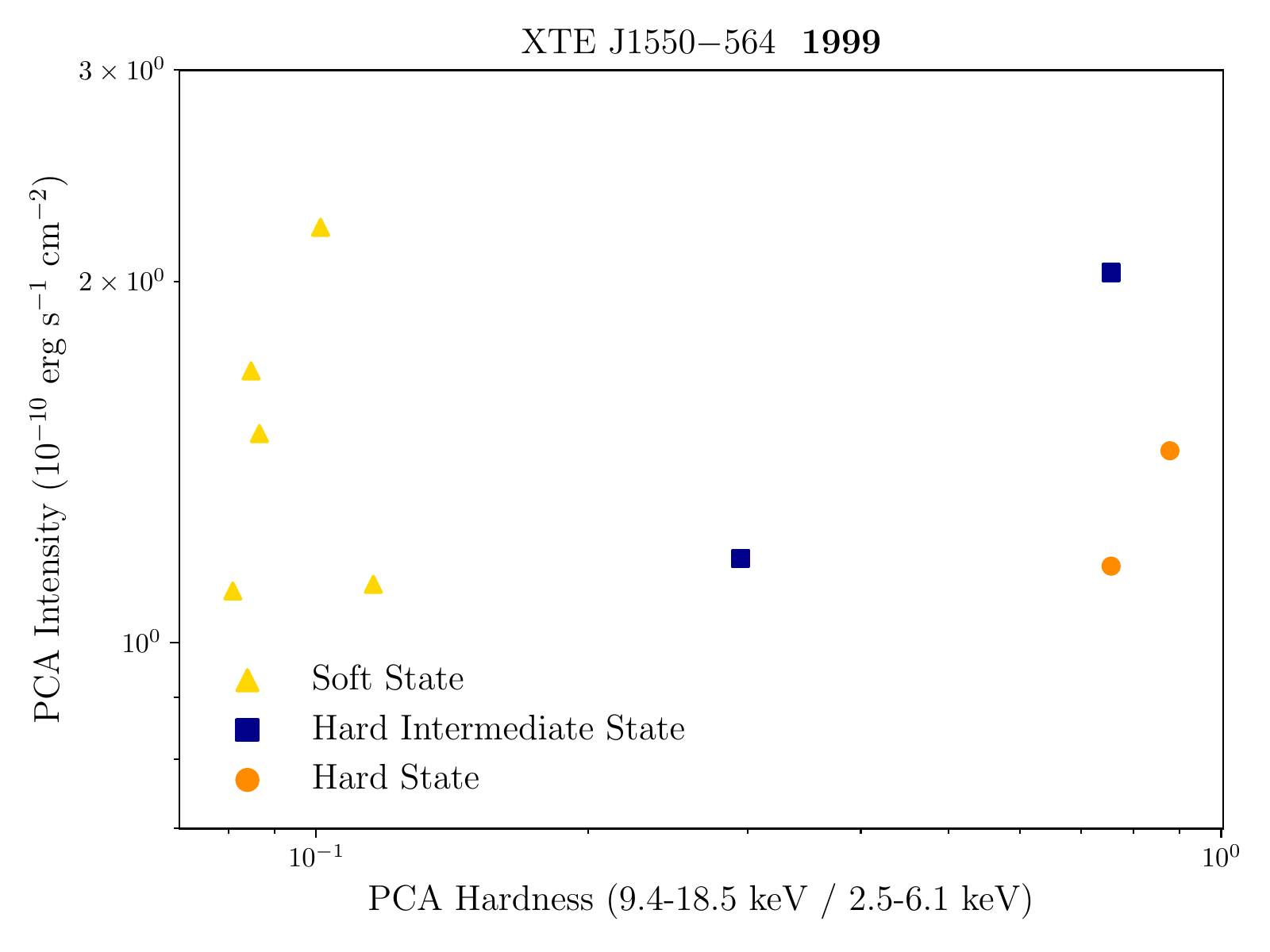}
\vspace{-0.50 cm}
\caption{\label{fig:XTE15501999_HID}
Hardness intensity diagram of XTE J1550$-$564 in 1999 outburst decay.}
\end{figure}

\begin{figure}
\includegraphics[width=85mm]{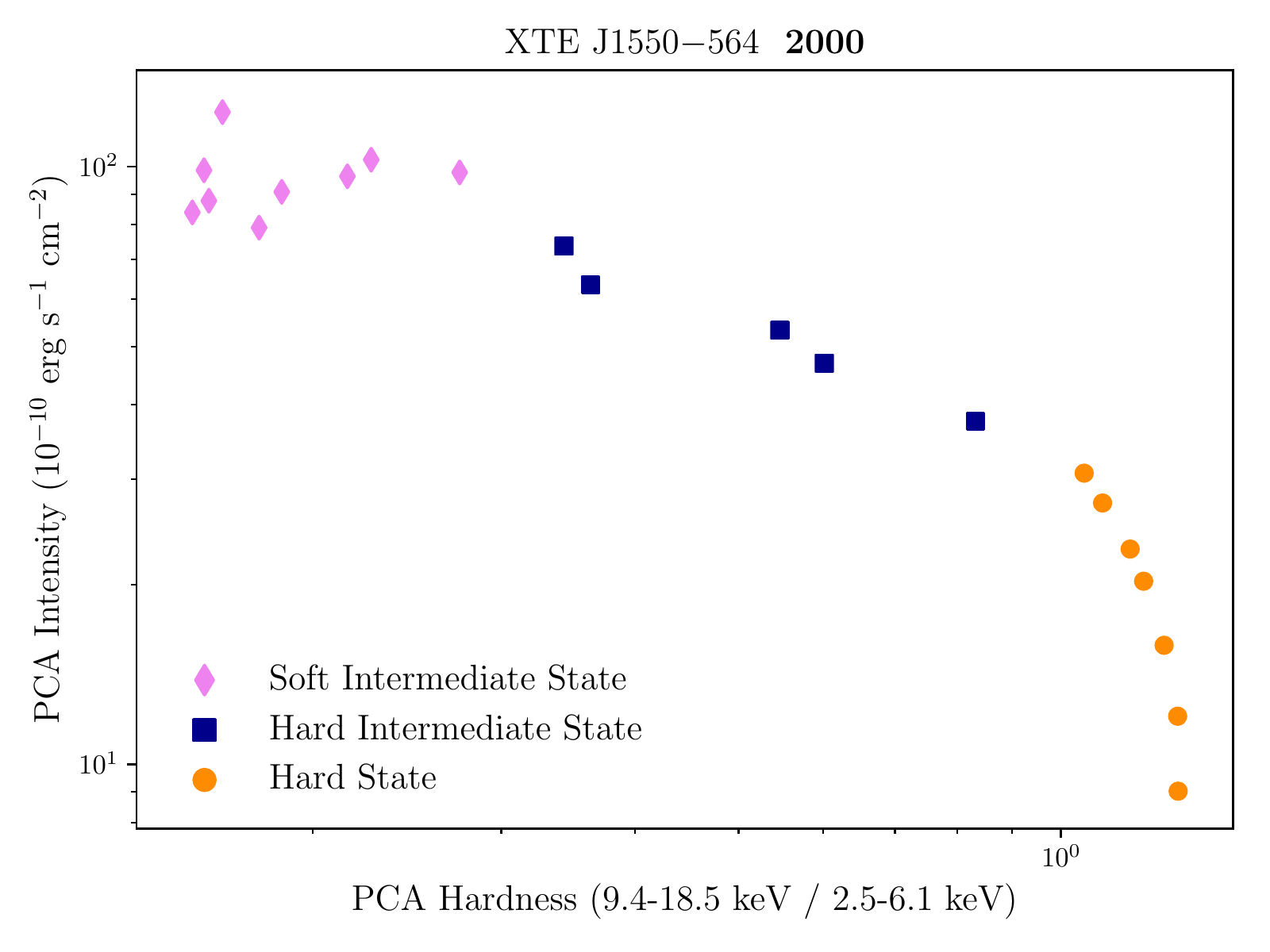}
\vspace{-0.50 cm}
\caption{\label{fig:XTE1550_2000}
Hardness intensity diagram of XTE J1550$-$564 in 2000 outburst decay.}
\end{figure}

\begin{figure}
\includegraphics[width=85mm]{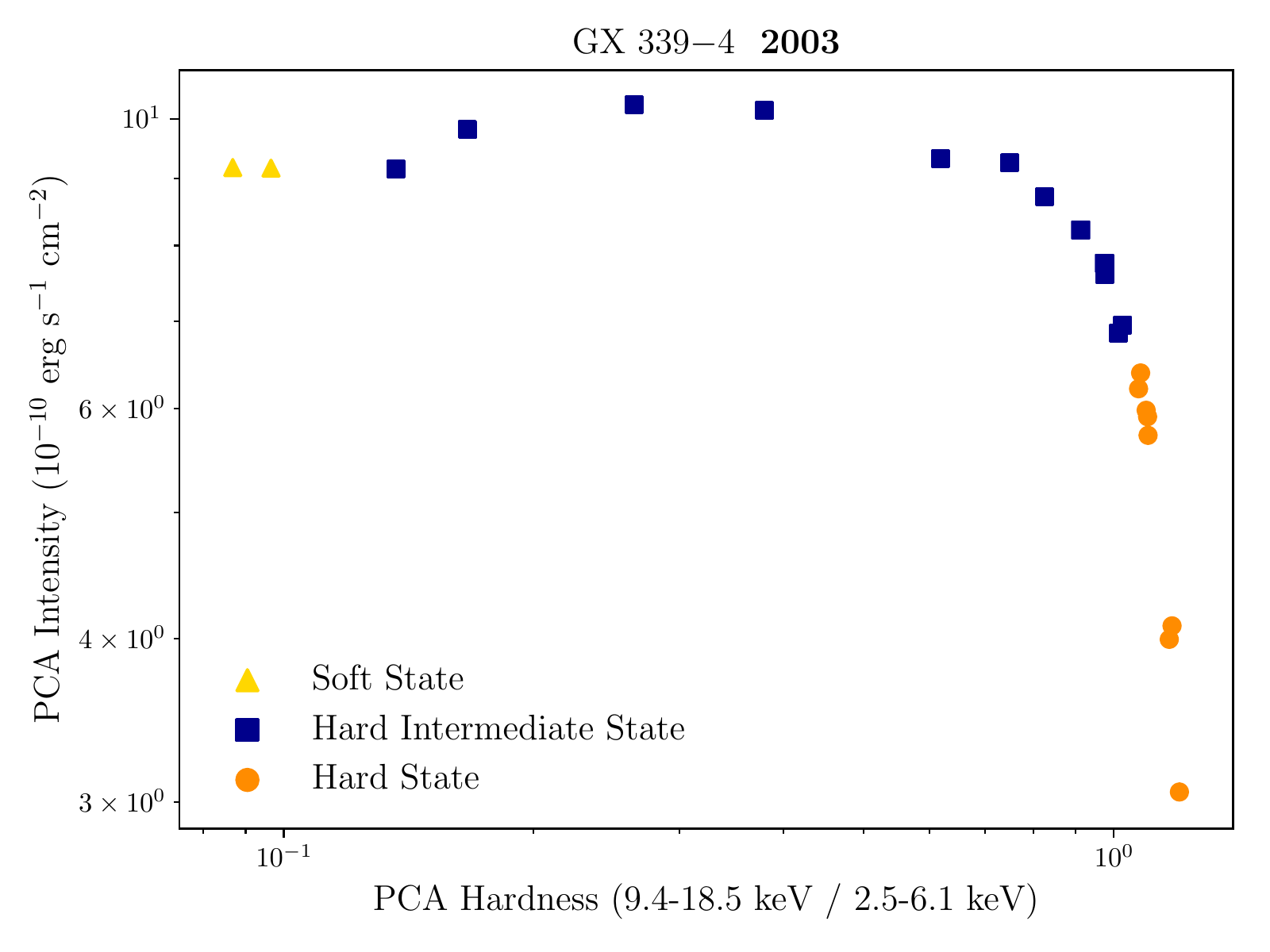}
\vspace{-0.50 cm}
\caption{\label{fig:GX2003_HID}
Hardness intensity diagram of GX 339$-$4 in 2003 outburst decay.}
\end{figure}

\begin{figure}
\includegraphics[width=85mm]{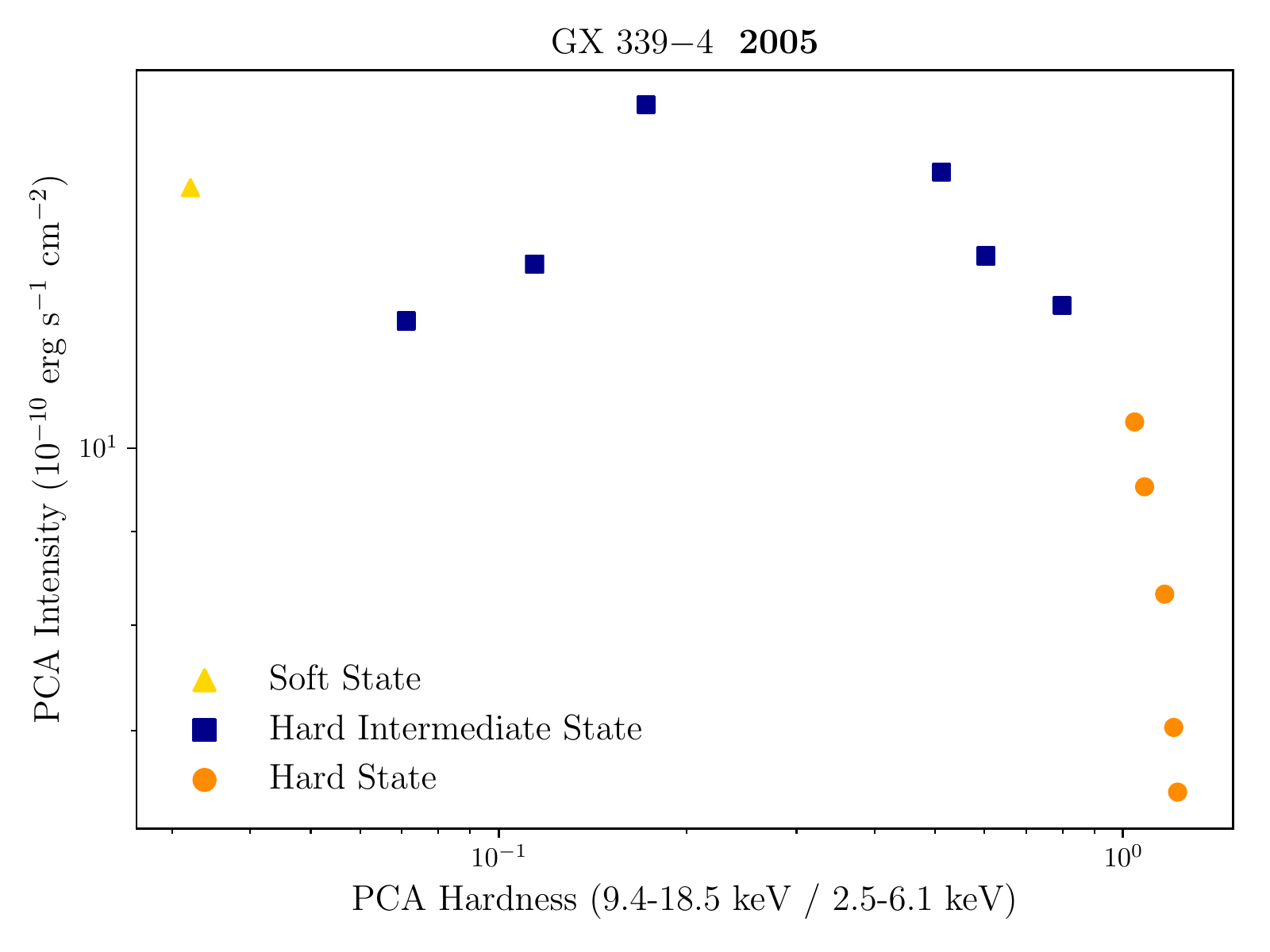}
\vspace{-0.50 cm}
\caption{\label{fig:GX2005_HID}
Hardness intensity diagram of GX 339$-$4 in 2005 outburst decay.}
\end{figure}

\begin{figure}
\includegraphics[width=85mm]{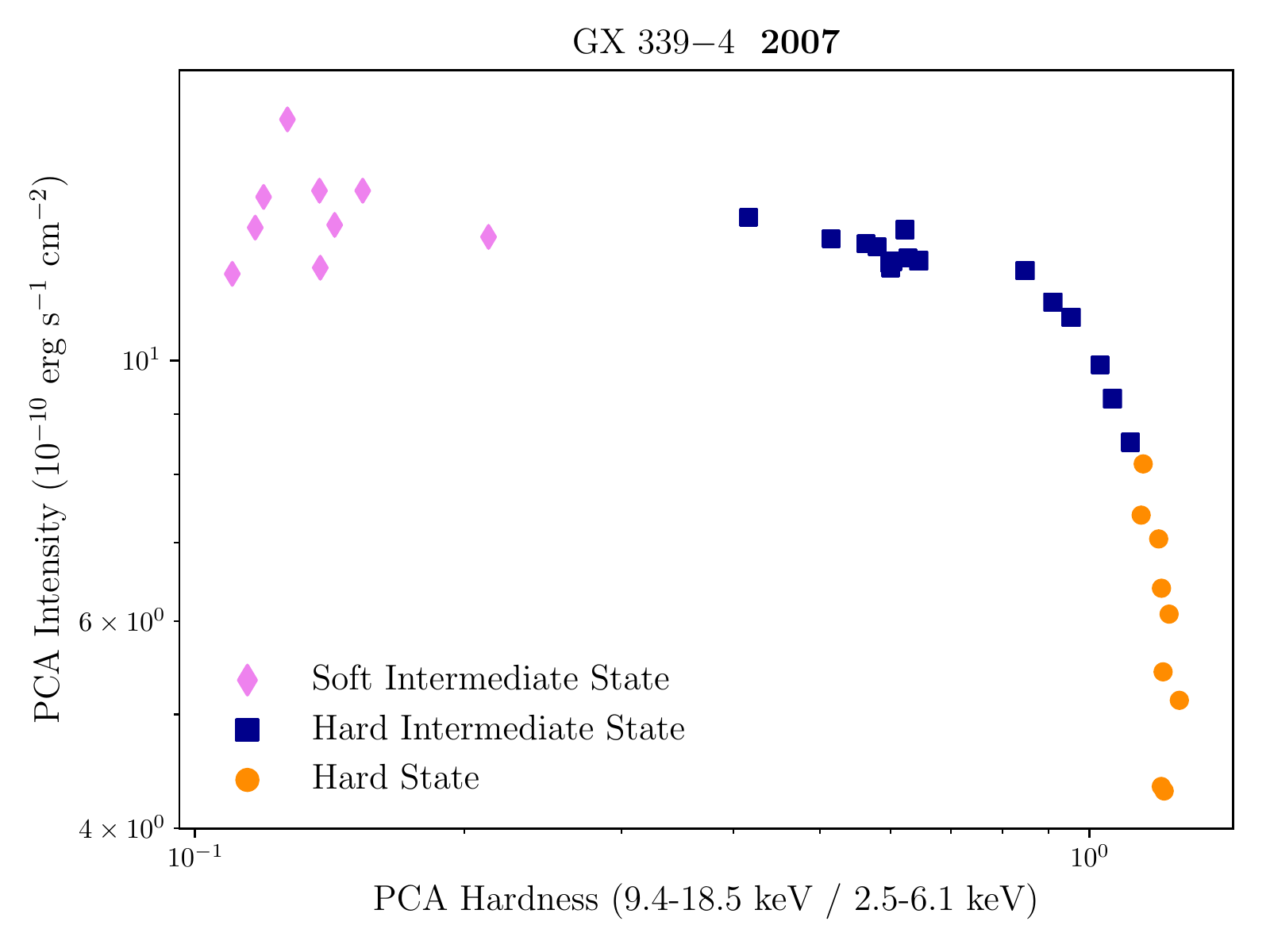}
\vspace{-0.50 cm}
\caption{\label{fig:GX2007_HID}
Hardness intensity diagram of GX 339$-$4 in 2007 outburst decay.}
\end{figure}

\begin{figure}
\includegraphics[width=85mm]{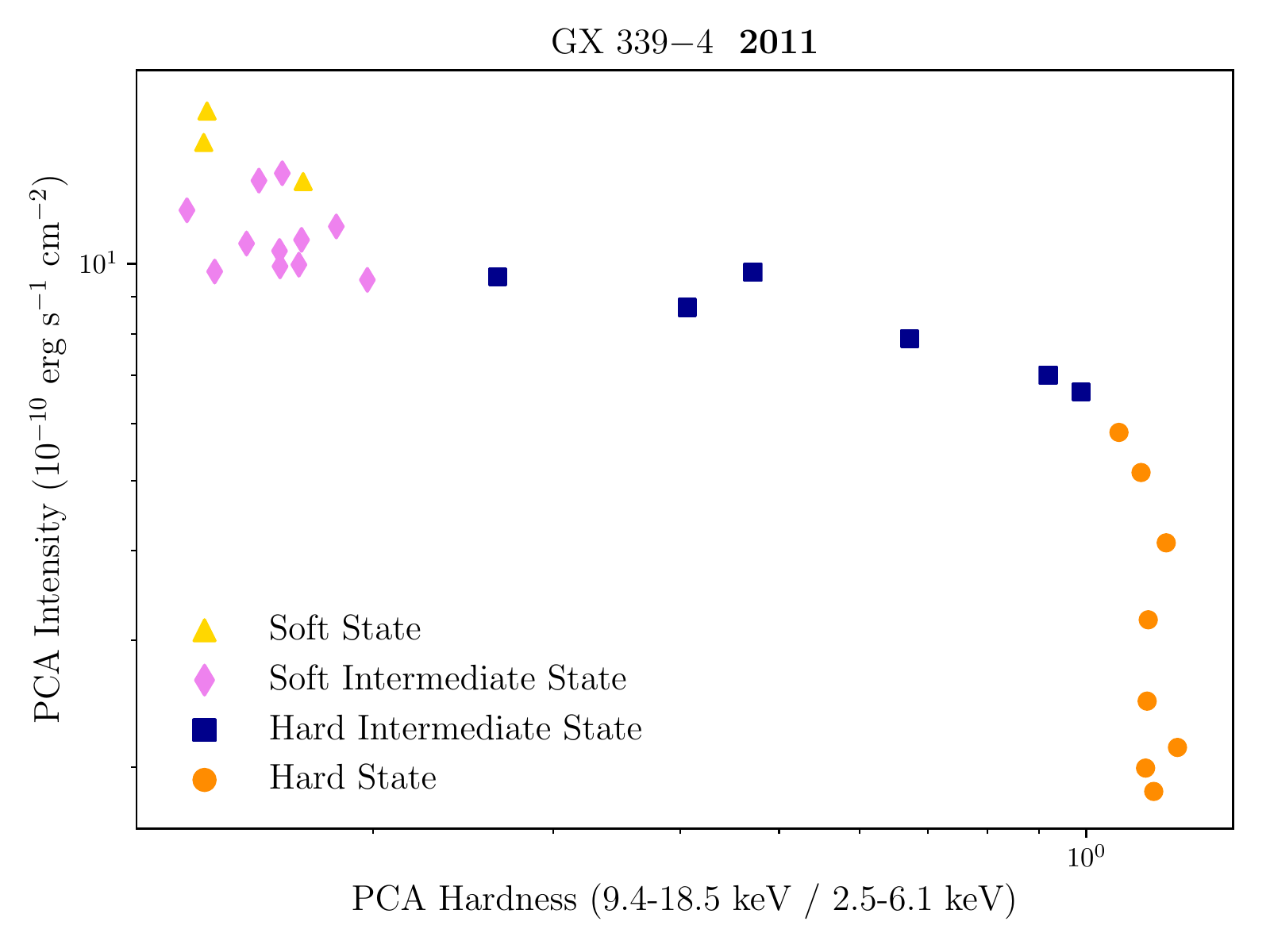}
\vspace{-0.50 cm}
\caption{\label{fig:GX2011_HID}
Hardness intensity diagram of GX 339$-$4 in 2011 outburst decay.}
\end{figure}

\begin{figure}
\includegraphics[width=85mm]{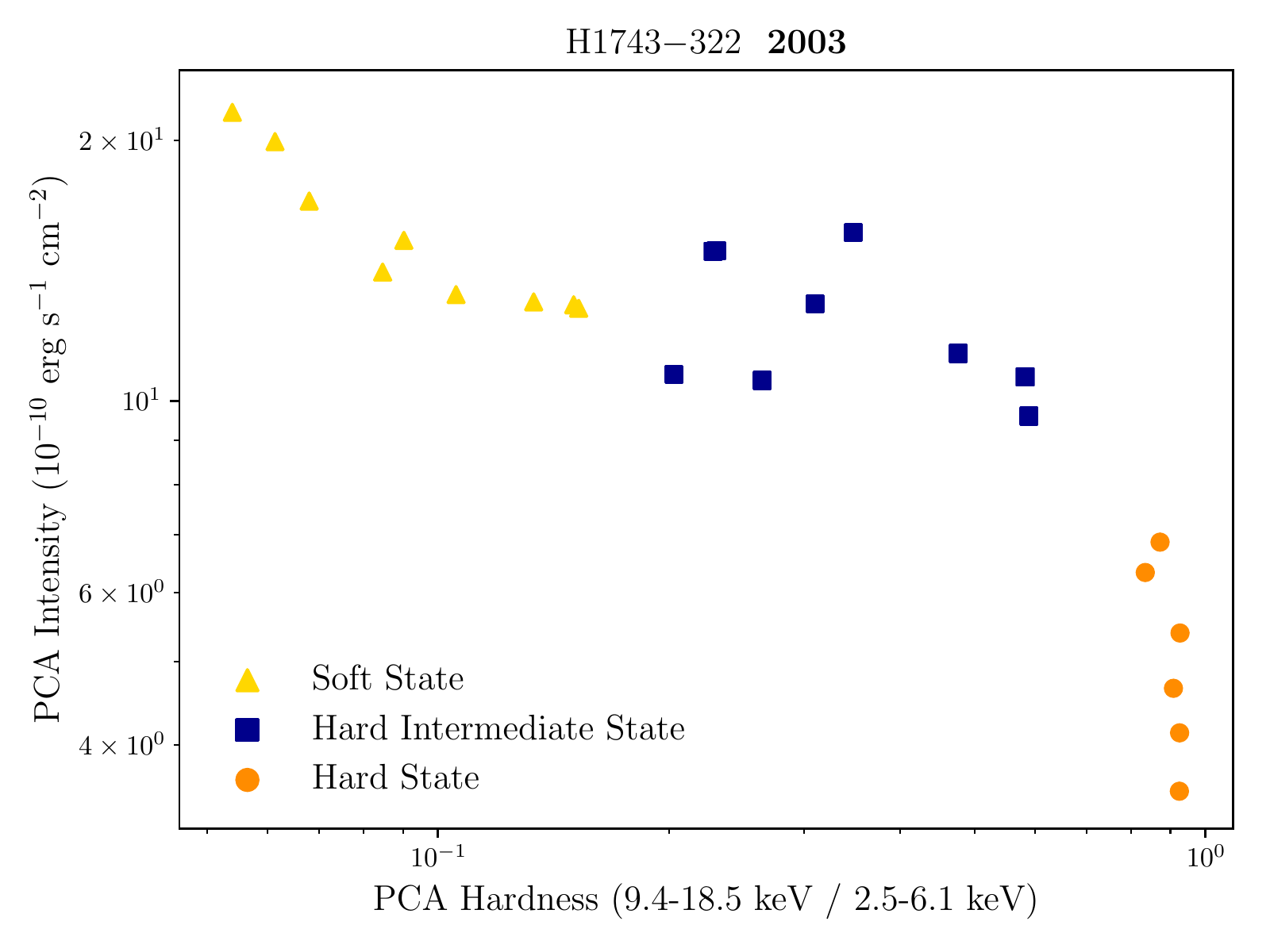}
\vspace{-0.50 cm}
\caption{\label{fig:H17432003_HID}
Hardness intensity diagram of H1743$-$322 in 2003 outburst decay.}
\end{figure}

\begin{figure}
\includegraphics[width=85mm]{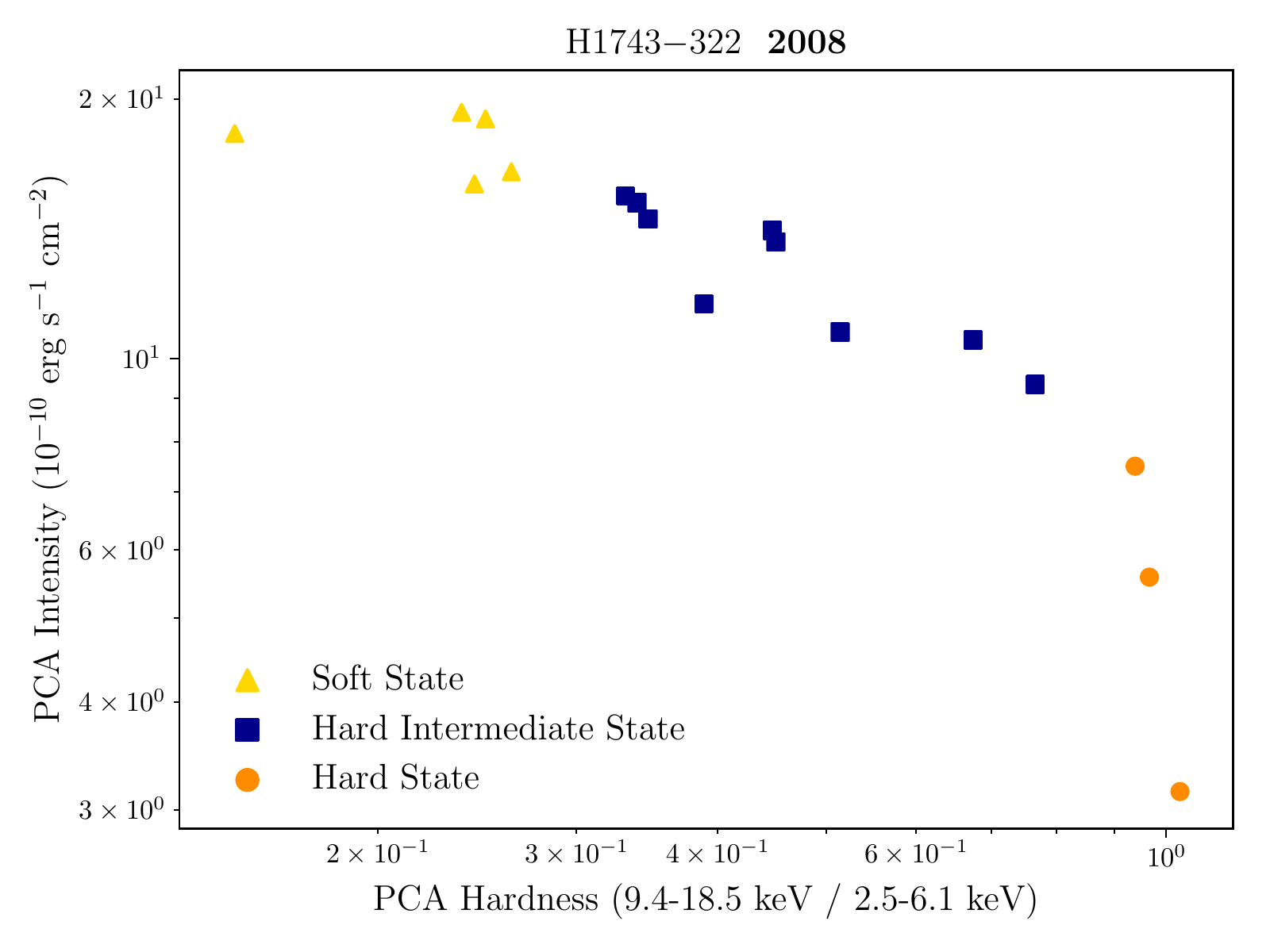}
\vspace{-0.50 cm}
\caption{\label{fig:H17432008_HID}
Hardness intensity diagram of H1743$-$322 in 2008 outburst decay.}
\end{figure}

\begin{figure}
\includegraphics[width=85mm]{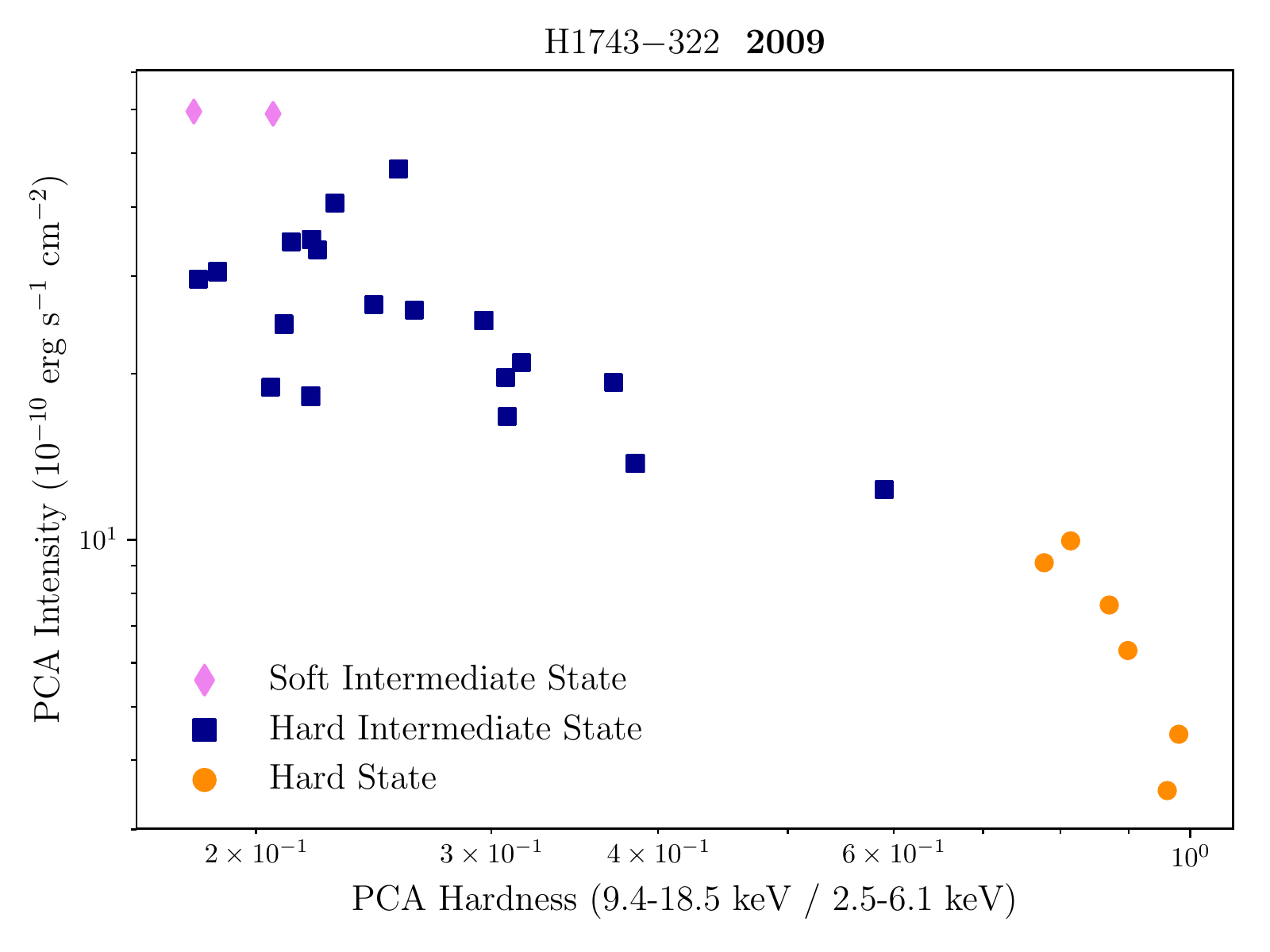}
\vspace{-0.50 cm}
\caption{\label{fig:H17432009_HID}
Hardness intensity diagram of H1743$-$322 in 2009 outburst decay.}
\end{figure}

\begin{figure}
\includegraphics[width=85mm]{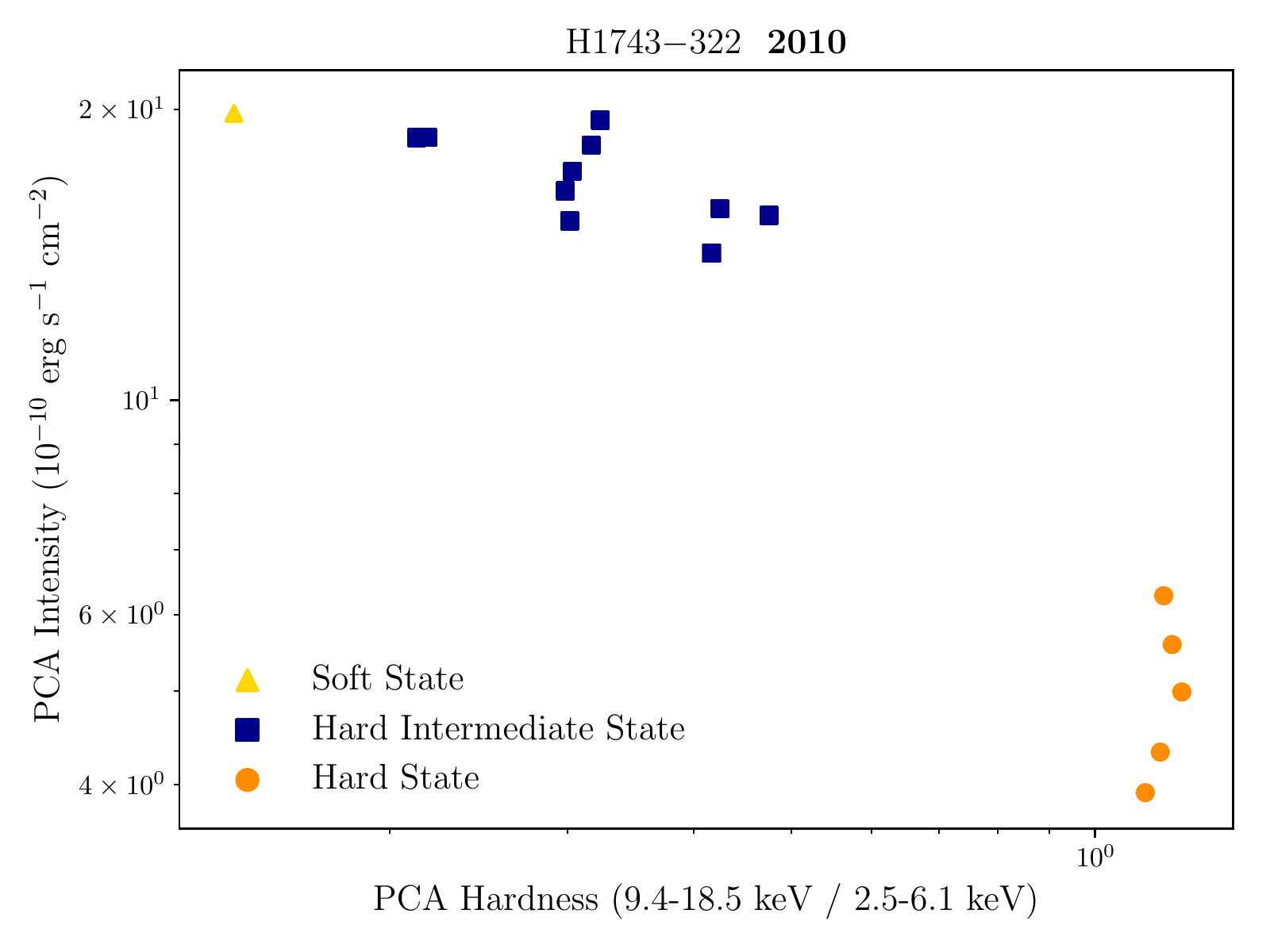}
\vspace{-0.50 cm}
\caption{\label{fig:H17432010_HID}
Hardness intensity diagram of H1743$-$322 in 2010 outburst decay.}
\end{figure}

\begin{figure}
\includegraphics[width=85mm]{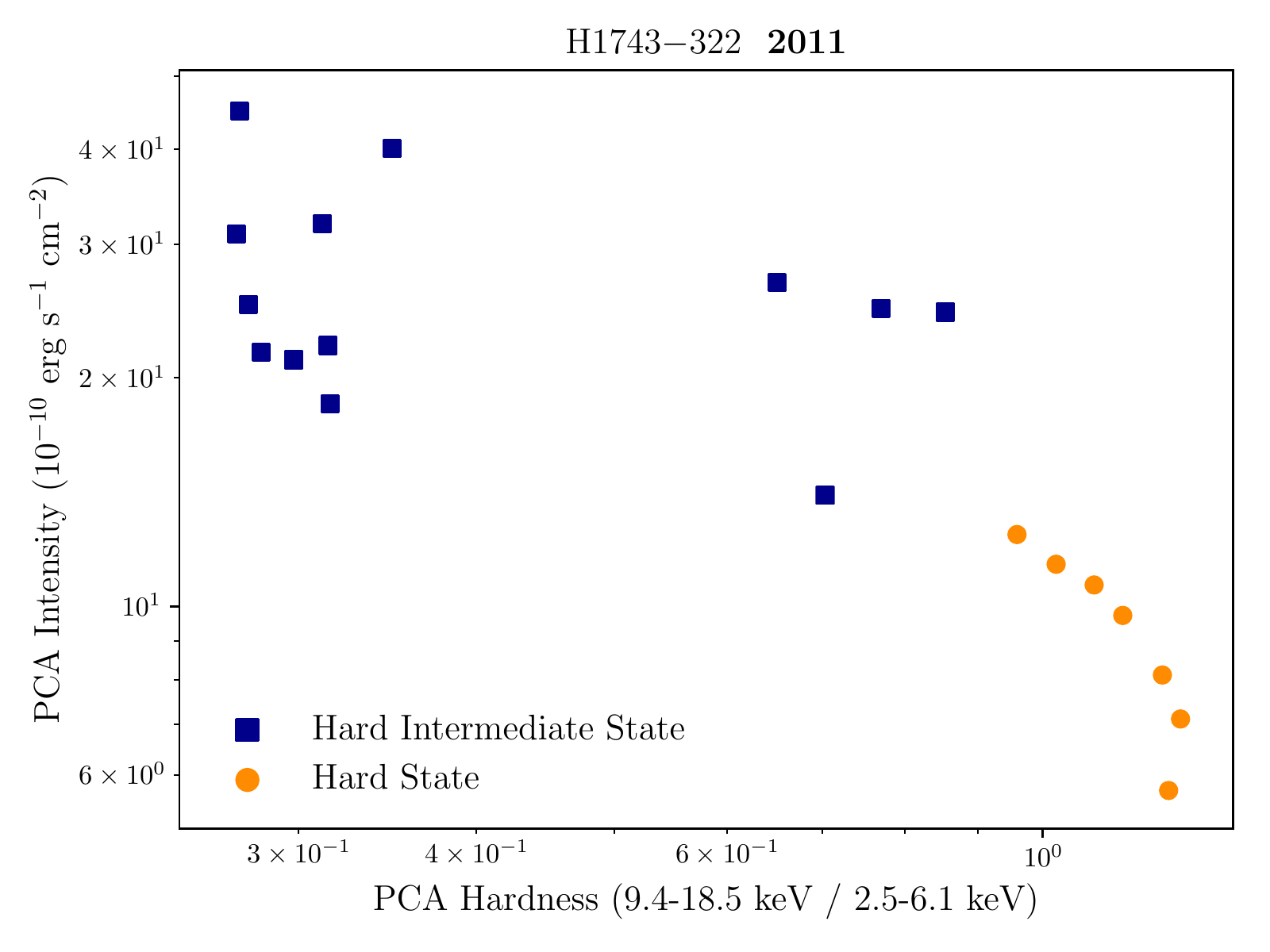}
\vspace{-0.50 cm}
\caption{\label{fig:H17432011_HID}
Hardness intensity diagram of H1743$-$322 in 2011 outburst decay.}
\end{figure}

\begin{figure}
\includegraphics[width=85mm]{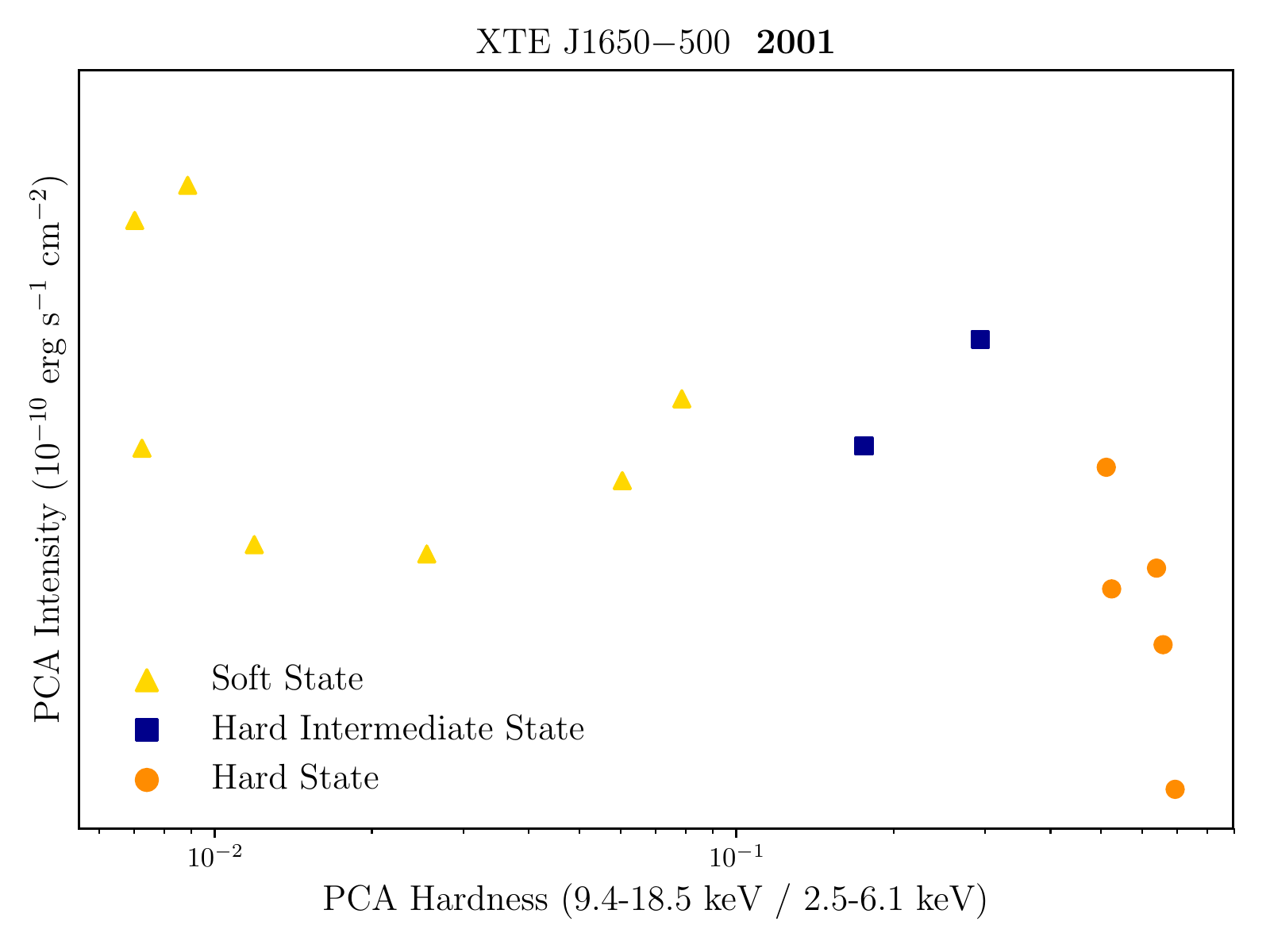}
\vspace{-0.50 cm}
\caption{\label{fig:J16502001_HID}
Hardness intensity diagram of XTE J1650$-$500 in 2001 outburst decay.}
\end{figure}

\begin{figure}
\includegraphics[width=85mm]{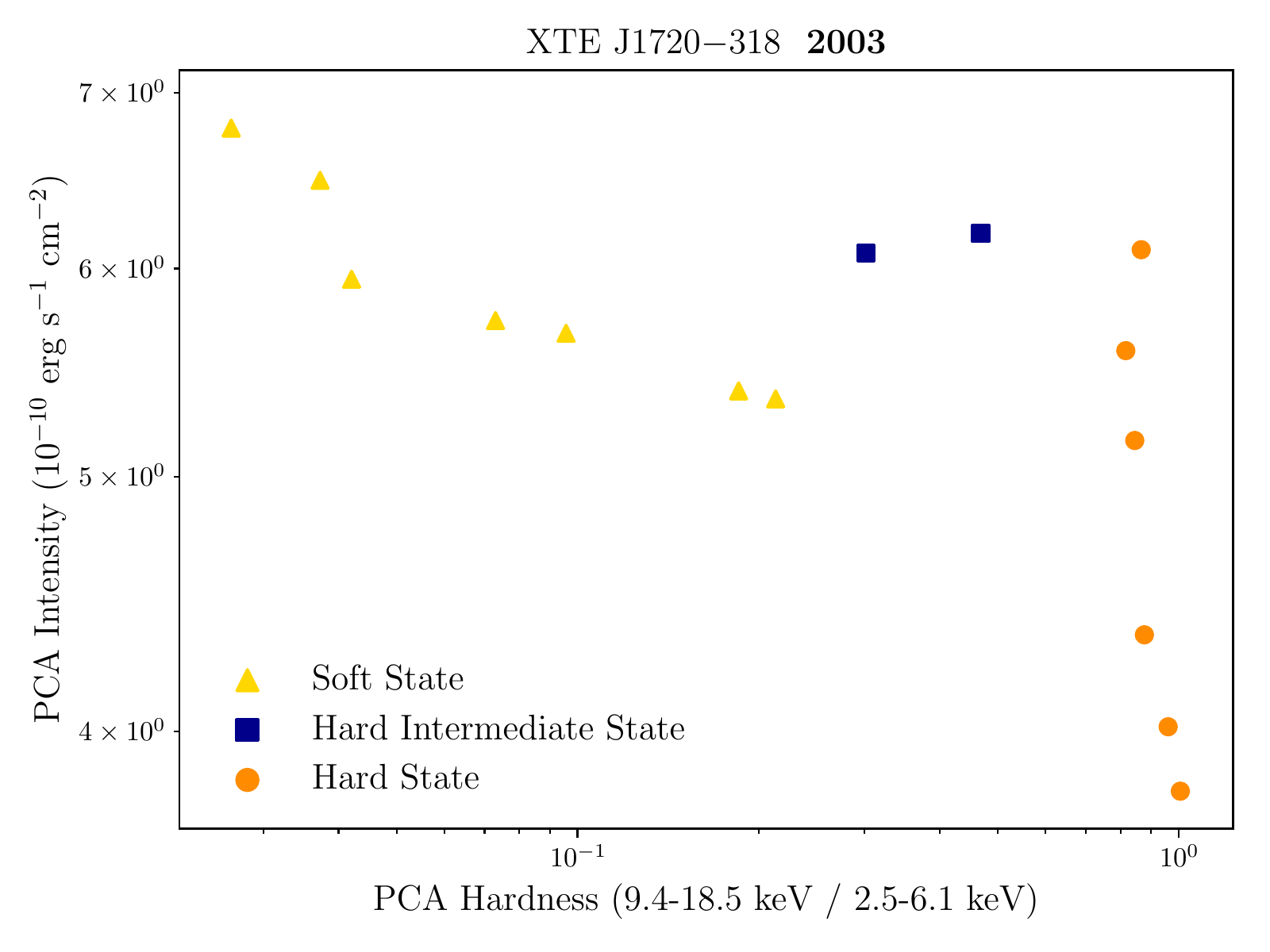}
\vspace{-0.50 cm}
\caption{\label{fig:J17202003_HID}
Hardness intensity diagram of XTE J1720$-$318 in 2003 outburst decay.}
\end{figure}

\begin{figure}
\includegraphics[width=85mm]{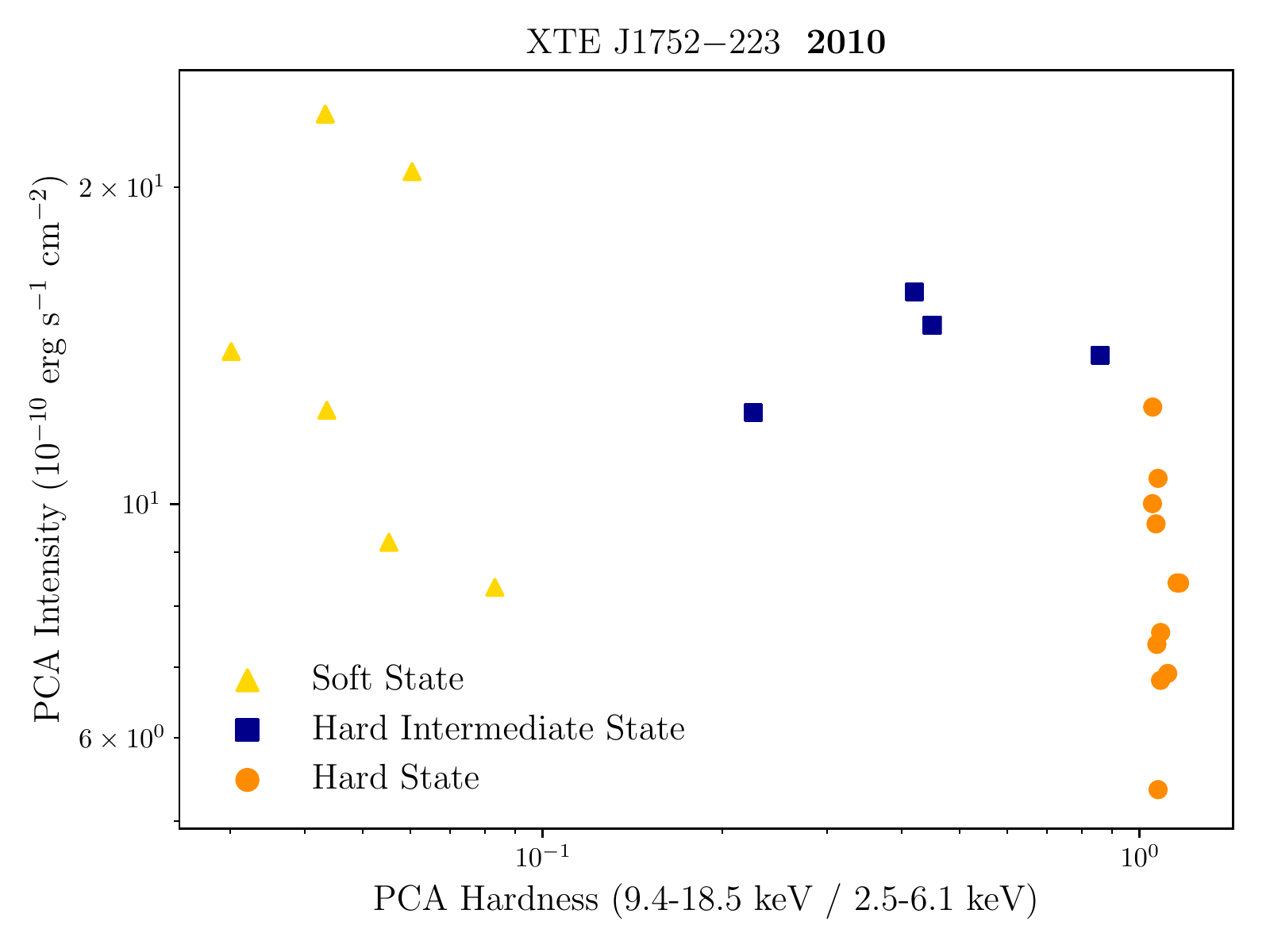}
\vspace{-0.50 cm}
\caption{\label{fig:J17522010_HID}
Hardness intensity diagram of XTE J1752$-$223 in 2010 outburst decay.}
\end{figure}

\begin{figure}
\includegraphics[width=85mm]{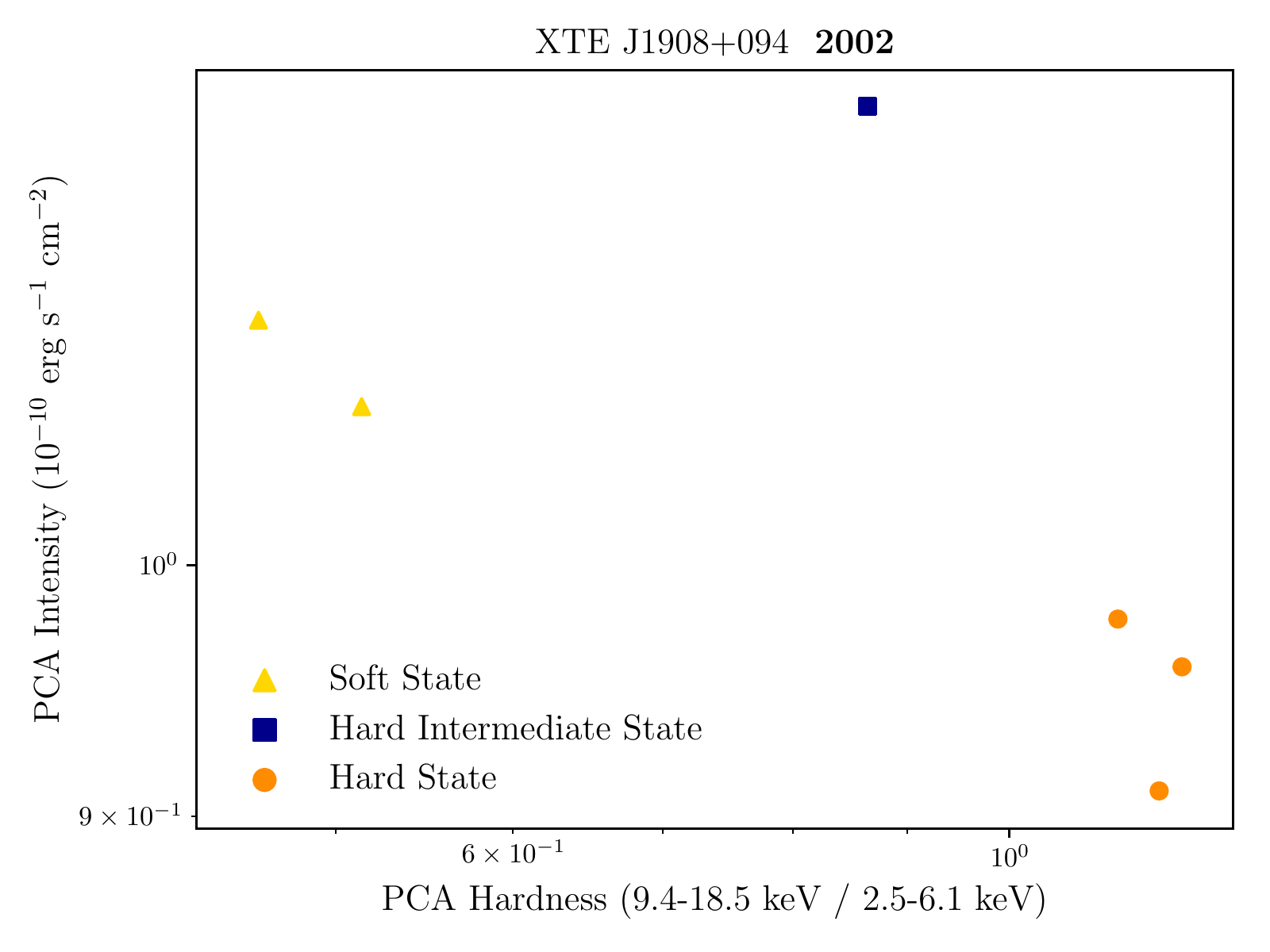}
\vspace{-0.50 cm}
\caption{\label{fig:J19082002_HID}
Hardness intensity diagram of XTE J1908+094 in 2002 outburst decay.}
\end{figure}

\begin{figure}
\includegraphics[width=85mm]{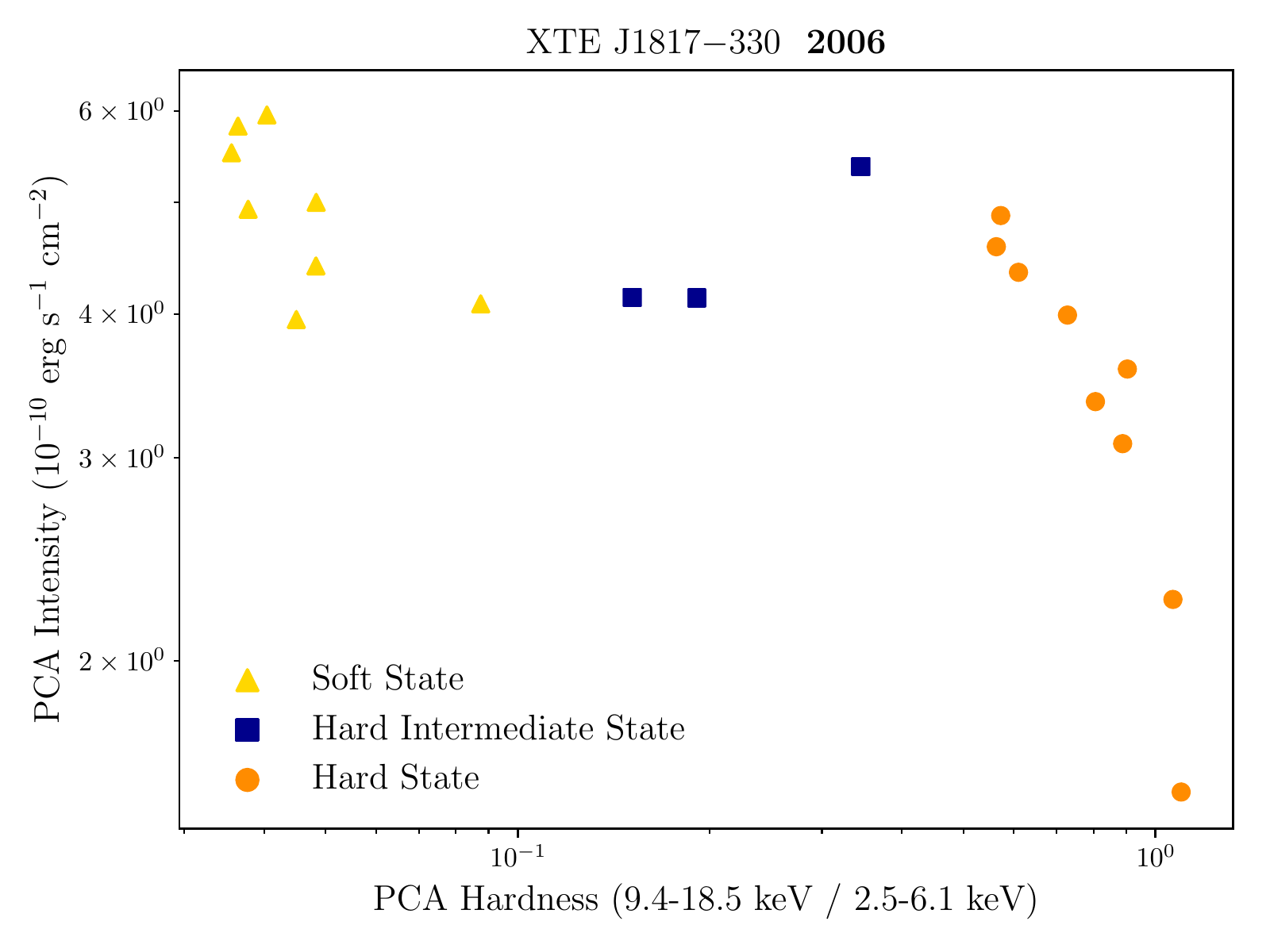}
\vspace{-0.50 cm}
\caption{\label{fig:J18172006_HID}
Hardness intensity diagram of XTE J1817$-$330 in 2006 outburst decay.}
\end{figure}

% Don't change these lines
\bsp	% typesetting comment
\label{lastpage}

\end{document}